\documentclass[10pt,sigconf,letterpaper,dvipsnames,nonacm,balance=false]{acmartnew}

\usepackage{booktabs}
\usepackage[T1]{fontenc}
\usepackage{graphics}
\usepackage[utf8]{inputenc}
\usepackage{latexsym}      % make some extra symbols available
\usepackage{lmodern}
\usepackage{microtype}
\microtypesetup{expansion=alltext,step=1} % hence we can activate expansion ...
\usepackage{multirow} 		% Multirow / multicolumn support 
\usepackage[group-digits=true, group-separator={,},scientific-notation = engineering]{siunitx}
\DeclareSIUnit{\nothing}{\relax}

\usepackage{tabularx}      % tables that auto-size
\usepackage{textcomp}
\usepackage[flushleft]{threeparttable}
\usepackage{xcolor}
\usepackage{xspace}		   % Fixes spacing problems in commands
\xspaceaddexceptions{\%}
\usepackage{natbib}

\graphicspath{{./}{./figures/}{./RFigures/}}

\newif\ifstatus
\newcommand{\todo}[1]{%
\ifstatus
  \begin{center}%
    \fbox{%
      \begin{minipage}{0.95\columnwidth}\small
        \textcolor{red}{\textbf{TODO:} #1}
      \end{minipage}%
    }
  \end{center}
\fi
}

\newcommand{\latinlocution}[1]{\textit{#1}}

\newcommand{\eg}{\latinlocution{e.g.,}\xspace}
\newcommand{\ie}{\latinlocution{i.e.,}\xspace}
\newcommand{\etal}{\latinlocution{et al.}\xspace}

\newcommand{\xref}[1]{\hyperref[#1]{\S\ref*{#1}}\xspace}


\newcommand{\first}{\emph{(i)}~}
\newcommand{\second}{\emph{(ii)}~}

\newcommand{\fref}[1]{Figure~\ref{#1}}
\newcommand{\tref}[1]{Table~\ref{#1}}
\newcommand{\cno}[0]{\textit{com}/\textit{net}/\textit{org}\xspace}
\newcommand{\dwnby}[1]{\textcolor{red}{$\downarrow$1}\xspace}

\newcommand{\tg}[1]{\textcolor{ForestGreen}{#1}\xspace}
\newcommand{\tgb}[1]{\textcolor{ForestGreen}{\textbf{#1}}\xspace}
\newcommand{\tgs}[1]{\textcolor{ForestGreen}{#1}}
\newcommand{\tgbs}[1]{\textcolor{ForestGreen}{\textbf{#1}}}

\newcommand{\trb}[1]{\textcolor{red}{\textbf{#1}}\xspace}
\newcommand{\tld}[1]{\textit{#1}\xspace}

\title[The Era of TLS 1.3]{The Era of TLS 1.3: Measuring Deployment and Use with Active and Passive Methods}
\author{Ralph Holz\textsuperscript{1}, Johanna Amann\textsuperscript{2}, Abbas Razaghpanah\textsuperscript{3}, and Narseo~Vallina-Rodriguez\textsuperscript{4}}

\affiliation{\textsuperscript{1}The University of Sydney, \textsuperscript{2}ICSI/Corelight/LBNL, \textsuperscript{3}Stony Brook University,\\ \textsuperscript{4}IMDEA Networks Institute/ICSI}

\newcommand{\ALLCONNSexact}[1]{301,390,779,750\xspace}
\newcommand{\ALLCONNS}[1]{301.4G\xspace}
\newcommand{\HTTPSCONNSexact}[1]{279,940,124,083\xspace}
\newcommand{\HTTPSCONNS}[1]{279.9G\xspace}
\newcommand{\VERSIONPREPARATIONexact}[1]{0\xspace}
\newcommand{\VERSIONPREPARATION}[1]{0\xspace}
\newcommand{\CIPHERPREPARATIONexact}[1]{0\xspace}
\newcommand{\CIPHERPREPARATION}[1]{0\xspace}
\newcommand{\negotiatedversionscategory}[1]{%
\ifnum\pdfstrcmp{#1}{TLSv10}=0 1.8\xspace\else\ifnum\pdfstrcmp{#1}{TLSv11}=0 0\xspace\else\ifnum\pdfstrcmp{#1}{TLSv12}=0 93.6\xspace\else\ifnum\pdfstrcmp{#1}{TLSv13}=0 4.6\xspace\else %
\textbf{XXX}\xspace %
\fi\fi\fi\fi%
}%
\newcommand{\signaturetablepreparationexact}[1]{0\xspace}
\newcommand{\signaturetablepreparation}[1]{0\xspace}
\newcommand{\SIGNATURETABLETOTALCONNSexact}[1]{1,739,785,078\xspace}
\newcommand{\SIGNATURETABLETOTALCONNS}[1]{1.7G\xspace}

\expandafter\def\csname storeCLIENTOFFEREDVERSIONS-1-1-\endcsname{TLSv12}
\expandafter\def\csname storeCLIENTOFFEREDVERSIONS-2-1-\endcsname{1.47G}
\expandafter\def\csname storeCLIENTOFFEREDVERSIONS-3-1-\endcsname{84.69}
\expandafter\def\csname storeCLIENTOFFEREDVERSIONS-1-2-\endcsname{TLSv13}
\expandafter\def\csname storeCLIENTOFFEREDVERSIONS-2-2-\endcsname{598.51M}
\expandafter\def\csname storeCLIENTOFFEREDVERSIONS-3-2-\endcsname{34.4}
\expandafter\def\csname storeCLIENTOFFEREDVERSIONS-1-3-\endcsname{TLSv10}
\expandafter\def\csname storeCLIENTOFFEREDVERSIONS-2-3-\endcsname{579.9M}
\expandafter\def\csname storeCLIENTOFFEREDVERSIONS-3-3-\endcsname{33.33}
\expandafter\def\csname storeCLIENTOFFEREDVERSIONS-1-4-\endcsname{TLSv11}
\expandafter\def\csname storeCLIENTOFFEREDVERSIONS-2-4-\endcsname{566.17M}
\expandafter\def\csname storeCLIENTOFFEREDVERSIONS-3-4-\endcsname{32.54}
\expandafter\def\csname storeCLIENTOFFEREDVERSIONS-1-5-\endcsname{TLSv13-FB26}
\expandafter\def\csname storeCLIENTOFFEREDVERSIONS-2-5-\endcsname{35.31M}
\expandafter\def\csname storeCLIENTOFFEREDVERSIONS-3-5-\endcsname{2.03}
\expandafter\def\csname storeCLIENTOFFEREDVERSIONS-1-6-\endcsname{TLSv13-draft23}
\expandafter\def\csname storeCLIENTOFFEREDVERSIONS-2-6-\endcsname{6.21M}
\expandafter\def\csname storeCLIENTOFFEREDVERSIONS-3-6-\endcsname{0.36}
\expandafter\def\csname storeCLIENTOFFEREDVERSIONS-1-7-\endcsname{TLSv13-draft18}
\expandafter\def\csname storeCLIENTOFFEREDVERSIONS-2-7-\endcsname{628.43K}
\expandafter\def\csname storeCLIENTOFFEREDVERSIONS-3-7-\endcsname{0.04}
\expandafter\def\csname storeCLIENTOFFEREDVERSIONS-1-8-\endcsname{TLSv13-draft28}
\expandafter\def\csname storeCLIENTOFFEREDVERSIONS-2-8-\endcsname{417.91K}
\expandafter\def\csname storeCLIENTOFFEREDVERSIONS-3-8-\endcsname{0.02}
\expandafter\def\csname storeCLIENTOFFEREDVERSIONS-1-9-\endcsname{TLSv13-draft26}
\expandafter\def\csname storeCLIENTOFFEREDVERSIONS-2-9-\endcsname{255.93K}
\expandafter\def\csname storeCLIENTOFFEREDVERSIONS-3-9-\endcsname{0.01}
\expandafter\def\csname storeCLIENTOFFEREDVERSIONS-1-10-\endcsname{TLSv13-FB23}
\expandafter\def\csname storeCLIENTOFFEREDVERSIONS-2-10-\endcsname{254.77K}
\expandafter\def\csname storeCLIENTOFFEREDVERSIONS-3-10-\endcsname{0.01}
\expandafter\def\csname storeCLIENTOFFEREDVERSIONS-1-11-\endcsname{TLSv13-7E02}
\expandafter\def\csname storeCLIENTOFFEREDVERSIONS-2-11-\endcsname{16.5K}
\expandafter\def\csname storeCLIENTOFFEREDVERSIONS-3-11-\endcsname{0}
\expandafter\def\csname storeCLIENTOFFEREDVERSIONS-1-12-\endcsname{TLSv13-draft27}
\expandafter\def\csname storeCLIENTOFFEREDVERSIONS-2-12-\endcsname{1.11K}
\expandafter\def\csname storeCLIENTOFFEREDVERSIONS-3-12-\endcsname{0}
\expandafter\def\csname storeCLIENTOFFEREDVERSIONS-1-13-\endcsname{TLSv13-7E01}
\expandafter\def\csname storeCLIENTOFFEREDVERSIONS-2-13-\endcsname{3}
\expandafter\def\csname storeCLIENTOFFEREDVERSIONS-3-13-\endcsname{0}
\def\CLIENTOFFEREDVERSIONSlines#1#2{\ifcsname storeCLIENTOFFEREDVERSIONS-#1-#2-\endcsname\csname storeCLIENTOFFEREDVERSIONS-#1-#2-\endcsname\else\textbf{XxX}\fi}

\expandafter\def\csname storeclientserverversionstable-1-1-\endcsname{TLSv10}
\expandafter\def\csname storeclientserverversionstable-2-1-\endcsname{1.83\%}
\expandafter\def\csname storeclientserverversionstable-3-1-\endcsname{33.33\%}
\expandafter\def\csname storeclientserverversionstable-1-2-\endcsname{TLSv11}
\expandafter\def\csname storeclientserverversionstable-2-2-\endcsname{0.01\%}
\expandafter\def\csname storeclientserverversionstable-3-2-\endcsname{32.54\%}
\expandafter\def\csname storeclientserverversionstable-1-3-\endcsname{TLSv12}
\expandafter\def\csname storeclientserverversionstable-2-3-\endcsname{93.6\%}
\expandafter\def\csname storeclientserverversionstable-3-3-\endcsname{84.69\%}
\expandafter\def\csname storeclientserverversionstable-1-4-\endcsname{TLSv13}
\expandafter\def\csname storeclientserverversionstable-2-4-\endcsname{2.51\%}
\expandafter\def\csname storeclientserverversionstable-3-4-\endcsname{34.4\%}
\expandafter\def\csname storeclientserverversionstable-1-5-\endcsname{TLSv13-7E01}
\expandafter\def\csname storeclientserverversionstable-2-5-\endcsname{none}
\expandafter\def\csname storeclientserverversionstable-3-5-\endcsname{0\%}
\expandafter\def\csname storeclientserverversionstable-1-6-\endcsname{TLSv13-7E02}
\expandafter\def\csname storeclientserverversionstable-2-6-\endcsname{none}
\expandafter\def\csname storeclientserverversionstable-3-6-\endcsname{0\%}
\expandafter\def\csname storeclientserverversionstable-1-7-\endcsname{TLSv13-draft18}
\expandafter\def\csname storeclientserverversionstable-2-7-\endcsname{none}
\expandafter\def\csname storeclientserverversionstable-3-7-\endcsname{0.04\%}
\expandafter\def\csname storeclientserverversionstable-1-8-\endcsname{TLSv13-draft23}
\expandafter\def\csname storeclientserverversionstable-2-8-\endcsname{0.01\%}
\expandafter\def\csname storeclientserverversionstable-3-8-\endcsname{0.36\%}
\expandafter\def\csname storeclientserverversionstable-1-9-\endcsname{TLSv13-draft26}
\expandafter\def\csname storeclientserverversionstable-2-9-\endcsname{0\%}
\expandafter\def\csname storeclientserverversionstable-3-9-\endcsname{0.01\%}
\expandafter\def\csname storeclientserverversionstable-1-10-\endcsname{TLSv13-draft27}
\expandafter\def\csname storeclientserverversionstable-2-10-\endcsname{none}
\expandafter\def\csname storeclientserverversionstable-3-10-\endcsname{0\%}
\expandafter\def\csname storeclientserverversionstable-1-11-\endcsname{TLSv13-draft28}
\expandafter\def\csname storeclientserverversionstable-2-11-\endcsname{0\%}
\expandafter\def\csname storeclientserverversionstable-3-11-\endcsname{0.02\%}
\expandafter\def\csname storeclientserverversionstable-1-12-\endcsname{TLSv13-FB23}
\expandafter\def\csname storeclientserverversionstable-2-12-\endcsname{0\%}
\expandafter\def\csname storeclientserverversionstable-3-12-\endcsname{0.01\%}
\expandafter\def\csname storeclientserverversionstable-1-13-\endcsname{TLSv13-FB26}
\expandafter\def\csname storeclientserverversionstable-2-13-\endcsname{2.05\%}
\expandafter\def\csname storeclientserverversionstable-3-13-\endcsname{2.03\%}
\def\clientserverversionstablelines#1#2{\ifcsname storeclientserverversionstable-#1-#2-\endcsname\csname storeclientserverversionstable-#1-#2-\endcsname\else\textbf{XxX}\fi}

\expandafter\def\csname storetlsonethreeusetable-1-1-\endcsname{60.62\%}
\expandafter\def\csname storetlsonethreeusetable-2-1-\endcsname{1}
\expandafter\def\csname storetlsonethreeusetable-3-1-\endcsname{1.09\%}
\expandafter\def\csname storetlsonethreeusetable-4-1-\endcsname{5}
\expandafter\def\csname storetlsonethreeusetable-5-1-\endcsname{facebook}
\expandafter\def\csname storetlsonethreeusetable-1-2-\endcsname{20.41\%}
\expandafter\def\csname storetlsonethreeusetable-2-2-\endcsname{2}
\expandafter\def\csname storetlsonethreeusetable-3-2-\endcsname{19.34\%}
\expandafter\def\csname storetlsonethreeusetable-4-2-\endcsname{2}
\expandafter\def\csname storetlsonethreeusetable-5-2-\endcsname{other}
\expandafter\def\csname storetlsonethreeusetable-1-3-\endcsname{9.66\%}
\expandafter\def\csname storetlsonethreeusetable-2-3-\endcsname{3}
\expandafter\def\csname storetlsonethreeusetable-3-3-\endcsname{70.45\%}
\expandafter\def\csname storetlsonethreeusetable-4-3-\endcsname{1}
\expandafter\def\csname storetlsonethreeusetable-5-3-\endcsname{cloudflare}
\expandafter\def\csname storetlsonethreeusetable-1-4-\endcsname{8.33\%}
\expandafter\def\csname storetlsonethreeusetable-2-4-\endcsname{4}
\expandafter\def\csname storetlsonethreeusetable-3-4-\endcsname{4.91\%}
\expandafter\def\csname storetlsonethreeusetable-4-4-\endcsname{3}
\expandafter\def\csname storetlsonethreeusetable-5-4-\endcsname{google}
\expandafter\def\csname storetlsonethreeusetable-1-5-\endcsname{0.42\%}
\expandafter\def\csname storetlsonethreeusetable-2-5-\endcsname{5}
\expandafter\def\csname storetlsonethreeusetable-3-5-\endcsname{2.32\%}
\expandafter\def\csname storetlsonethreeusetable-4-5-\endcsname{4}
\expandafter\def\csname storetlsonethreeusetable-5-5-\endcsname{amazon}
\expandafter\def\csname storetlsonethreeusetable-1-6-\endcsname{0.41\%}
\expandafter\def\csname storetlsonethreeusetable-2-6-\endcsname{6}
\expandafter\def\csname storetlsonethreeusetable-3-6-\endcsname{0.86\%}
\expandafter\def\csname storetlsonethreeusetable-4-6-\endcsname{6}
\expandafter\def\csname storetlsonethreeusetable-5-6-\endcsname{akamai}
\expandafter\def\csname storetlsonethreeusetable-1-7-\endcsname{0.05\%}
\expandafter\def\csname storetlsonethreeusetable-2-7-\endcsname{7}
\expandafter\def\csname storetlsonethreeusetable-3-7-\endcsname{0.65\%}
\expandafter\def\csname storetlsonethreeusetable-4-7-\endcsname{7}
\expandafter\def\csname storetlsonethreeusetable-5-7-\endcsname{digitalocean}
\expandafter\def\csname storetlsonethreeusetable-1-8-\endcsname{0.04\%}
\expandafter\def\csname storetlsonethreeusetable-2-8-\endcsname{8}
\expandafter\def\csname storetlsonethreeusetable-3-8-\endcsname{0.01\%}
\expandafter\def\csname storetlsonethreeusetable-4-8-\endcsname{11}
\expandafter\def\csname storetlsonethreeusetable-5-8-\endcsname{squarespace}
\expandafter\def\csname storetlsonethreeusetable-1-9-\endcsname{0.02\%}
\expandafter\def\csname storetlsonethreeusetable-2-9-\endcsname{9}
\expandafter\def\csname storetlsonethreeusetable-3-9-\endcsname{0.04\%}
\expandafter\def\csname storetlsonethreeusetable-4-9-\endcsname{10}
\expandafter\def\csname storetlsonethreeusetable-5-9-\endcsname{alibaba}
\expandafter\def\csname storetlsonethreeusetable-1-10-\endcsname{0.02\%}
\expandafter\def\csname storetlsonethreeusetable-2-10-\endcsname{10}
\expandafter\def\csname storetlsonethreeusetable-3-10-\endcsname{0.24\%}
\expandafter\def\csname storetlsonethreeusetable-4-10-\endcsname{8}
\expandafter\def\csname storetlsonethreeusetable-5-10-\endcsname{ovh}
\expandafter\def\csname storetlsonethreeusetable-1-11-\endcsname{0\%}
\expandafter\def\csname storetlsonethreeusetable-2-11-\endcsname{11}
\expandafter\def\csname storetlsonethreeusetable-3-11-\endcsname{0.07\%}
\expandafter\def\csname storetlsonethreeusetable-4-11-\endcsname{9}
\expandafter\def\csname storetlsonethreeusetable-5-11-\endcsname{azure}
\expandafter\def\csname storetlsonethreeusetable-1-12-\endcsname{0\%}
\expandafter\def\csname storetlsonethreeusetable-2-12-\endcsname{12}
\expandafter\def\csname storetlsonethreeusetable-3-12-\endcsname{0.01\%}
\expandafter\def\csname storetlsonethreeusetable-4-12-\endcsname{12}
\expandafter\def\csname storetlsonethreeusetable-5-12-\endcsname{godaddy}
\def\tlsonethreeusetablelines#1#2{\ifcsname storetlsonethreeusetable-#1-#2-\endcsname\csname storetlsonethreeusetable-#1-#2-\endcsname\else\textbf{XxX}\fi}

\expandafter\def\csname storetlsonethreeusetablebig-1-1-\endcsname{facebook}
\expandafter\def\csname storetlsonethreeusetablebig-2-1-\endcsname{60.78\%}
\expandafter\def\csname storetlsonethreeusetablebig-3-1-\endcsname{1}
\expandafter\def\csname storetlsonethreeusetablebig-4-1-\endcsname{1.02\%}
\expandafter\def\csname storetlsonethreeusetablebig-5-1-\endcsname{7}
\expandafter\def\csname storetlsonethreeusetablebig-6-1-\endcsname{3.22\%}
\expandafter\def\csname storetlsonethreeusetablebig-7-1-\endcsname{4}
\expandafter\def\csname storetlsonethreeusetablebig-8-1-\endcsname{0.12\%}
\expandafter\def\csname storetlsonethreeusetablebig-9-1-\endcsname{10}
\expandafter\def\csname storetlsonethreeusetablebig-1-2-\endcsname{other}
\expandafter\def\csname storetlsonethreeusetablebig-2-2-\endcsname{20.26\%}
\expandafter\def\csname storetlsonethreeusetablebig-3-2-\endcsname{2}
\expandafter\def\csname storetlsonethreeusetablebig-4-2-\endcsname{34.54\%}
\expandafter\def\csname storetlsonethreeusetablebig-5-2-\endcsname{1}
\expandafter\def\csname storetlsonethreeusetablebig-6-2-\endcsname{17.2\%}
\expandafter\def\csname storetlsonethreeusetablebig-7-2-\endcsname{2}
\expandafter\def\csname storetlsonethreeusetablebig-8-2-\endcsname{17.42\%}
\expandafter\def\csname storetlsonethreeusetablebig-9-2-\endcsname{2}
\expandafter\def\csname storetlsonethreeusetablebig-1-3-\endcsname{cloudflare}
\expandafter\def\csname storetlsonethreeusetablebig-2-3-\endcsname{9.66\%}
\expandafter\def\csname storetlsonethreeusetablebig-3-3-\endcsname{3}
\expandafter\def\csname storetlsonethreeusetablebig-4-3-\endcsname{1.39\%}
\expandafter\def\csname storetlsonethreeusetablebig-5-3-\endcsname{6}
\expandafter\def\csname storetlsonethreeusetablebig-6-3-\endcsname{70.45\%}
\expandafter\def\csname storetlsonethreeusetablebig-7-3-\endcsname{1}
\expandafter\def\csname storetlsonethreeusetablebig-8-3-\endcsname{4.86\%}
\expandafter\def\csname storetlsonethreeusetablebig-9-3-\endcsname{4}
\expandafter\def\csname storetlsonethreeusetablebig-1-4-\endcsname{google}
\expandafter\def\csname storetlsonethreeusetablebig-2-4-\endcsname{8.33\%}
\expandafter\def\csname storetlsonethreeusetablebig-3-4-\endcsname{4}
\expandafter\def\csname storetlsonethreeusetablebig-4-4-\endcsname{12.87\%}
\expandafter\def\csname storetlsonethreeusetablebig-5-4-\endcsname{3}
\expandafter\def\csname storetlsonethreeusetablebig-6-4-\endcsname{4.91\%}
\expandafter\def\csname storetlsonethreeusetablebig-7-4-\endcsname{3}
\expandafter\def\csname storetlsonethreeusetablebig-8-4-\endcsname{1.47\%}
\expandafter\def\csname storetlsonethreeusetablebig-9-4-\endcsname{5}
\expandafter\def\csname storetlsonethreeusetablebig-1-5-\endcsname{amazon}
\expandafter\def\csname storetlsonethreeusetablebig-2-5-\endcsname{0.42\%}
\expandafter\def\csname storetlsonethreeusetablebig-3-5-\endcsname{5}
\expandafter\def\csname storetlsonethreeusetablebig-4-5-\endcsname{33.64\%}
\expandafter\def\csname storetlsonethreeusetablebig-5-5-\endcsname{2}
\expandafter\def\csname storetlsonethreeusetablebig-6-5-\endcsname{2.32\%}
\expandafter\def\csname storetlsonethreeusetablebig-7-5-\endcsname{5}
\expandafter\def\csname storetlsonethreeusetablebig-8-5-\endcsname{68.02\%}
\expandafter\def\csname storetlsonethreeusetablebig-9-5-\endcsname{1}
\expandafter\def\csname storetlsonethreeusetablebig-1-6-\endcsname{akamai}
\expandafter\def\csname storetlsonethreeusetablebig-2-6-\endcsname{0.41\%}
\expandafter\def\csname storetlsonethreeusetablebig-3-6-\endcsname{6}
\expandafter\def\csname storetlsonethreeusetablebig-4-6-\endcsname{7.2\%}
\expandafter\def\csname storetlsonethreeusetablebig-5-6-\endcsname{5}
\expandafter\def\csname storetlsonethreeusetablebig-6-6-\endcsname{0.86\%}
\expandafter\def\csname storetlsonethreeusetablebig-7-6-\endcsname{6}
\expandafter\def\csname storetlsonethreeusetablebig-8-6-\endcsname{6.13\%}
\expandafter\def\csname storetlsonethreeusetablebig-9-6-\endcsname{3}
\expandafter\def\csname storetlsonethreeusetablebig-1-7-\endcsname{digitalocean}
\expandafter\def\csname storetlsonethreeusetablebig-2-7-\endcsname{0.05\%}
\expandafter\def\csname storetlsonethreeusetablebig-3-7-\endcsname{7}
\expandafter\def\csname storetlsonethreeusetablebig-4-7-\endcsname{0.16\%}
\expandafter\def\csname storetlsonethreeusetablebig-5-7-\endcsname{9}
\expandafter\def\csname storetlsonethreeusetablebig-6-7-\endcsname{0.65\%}
\expandafter\def\csname storetlsonethreeusetablebig-7-7-\endcsname{7}
\expandafter\def\csname storetlsonethreeusetablebig-8-7-\endcsname{0.69\%}
\expandafter\def\csname storetlsonethreeusetablebig-9-7-\endcsname{6}
\expandafter\def\csname storetlsonethreeusetablebig-1-8-\endcsname{squarespace}
\expandafter\def\csname storetlsonethreeusetablebig-2-8-\endcsname{0.04\%}
\expandafter\def\csname storetlsonethreeusetablebig-3-8-\endcsname{8}
\expandafter\def\csname storetlsonethreeusetablebig-4-8-\endcsname{0\%}
\expandafter\def\csname storetlsonethreeusetablebig-5-8-\endcsname{12}
\expandafter\def\csname storetlsonethreeusetablebig-6-8-\endcsname{0.01\%}
\expandafter\def\csname storetlsonethreeusetablebig-7-8-\endcsname{11}
\expandafter\def\csname storetlsonethreeusetablebig-8-8-\endcsname{0\%}
\expandafter\def\csname storetlsonethreeusetablebig-9-8-\endcsname{12}
\expandafter\def\csname storetlsonethreeusetablebig-1-9-\endcsname{alibaba}
\expandafter\def\csname storetlsonethreeusetablebig-2-9-\endcsname{0.02\%}
\expandafter\def\csname storetlsonethreeusetablebig-3-9-\endcsname{9}
\expandafter\def\csname storetlsonethreeusetablebig-4-9-\endcsname{0.04\%}
\expandafter\def\csname storetlsonethreeusetablebig-5-9-\endcsname{11}
\expandafter\def\csname storetlsonethreeusetablebig-6-9-\endcsname{0.04\%}
\expandafter\def\csname storetlsonethreeusetablebig-7-9-\endcsname{10}
\expandafter\def\csname storetlsonethreeusetablebig-8-9-\endcsname{0.04\%}
\expandafter\def\csname storetlsonethreeusetablebig-9-9-\endcsname{11}
\expandafter\def\csname storetlsonethreeusetablebig-1-10-\endcsname{ovh}
\expandafter\def\csname storetlsonethreeusetablebig-2-10-\endcsname{0.02\%}
\expandafter\def\csname storetlsonethreeusetablebig-3-10-\endcsname{10}
\expandafter\def\csname storetlsonethreeusetablebig-4-10-\endcsname{0.2\%}
\expandafter\def\csname storetlsonethreeusetablebig-5-10-\endcsname{8}
\expandafter\def\csname storetlsonethreeusetablebig-6-10-\endcsname{0.24\%}
\expandafter\def\csname storetlsonethreeusetablebig-7-10-\endcsname{8}
\expandafter\def\csname storetlsonethreeusetablebig-8-10-\endcsname{0.43\%}
\expandafter\def\csname storetlsonethreeusetablebig-9-10-\endcsname{8}
\expandafter\def\csname storetlsonethreeusetablebig-1-11-\endcsname{azure}
\expandafter\def\csname storetlsonethreeusetablebig-2-11-\endcsname{0\%}
\expandafter\def\csname storetlsonethreeusetablebig-3-11-\endcsname{11}
\expandafter\def\csname storetlsonethreeusetablebig-4-11-\endcsname{8.88\%}
\expandafter\def\csname storetlsonethreeusetablebig-5-11-\endcsname{4}
\expandafter\def\csname storetlsonethreeusetablebig-6-11-\endcsname{0.07\%}
\expandafter\def\csname storetlsonethreeusetablebig-7-11-\endcsname{9}
\expandafter\def\csname storetlsonethreeusetablebig-8-11-\endcsname{0.45\%}
\expandafter\def\csname storetlsonethreeusetablebig-9-11-\endcsname{7}
\expandafter\def\csname storetlsonethreeusetablebig-1-12-\endcsname{godaddy}
\expandafter\def\csname storetlsonethreeusetablebig-2-12-\endcsname{0\%}
\expandafter\def\csname storetlsonethreeusetablebig-3-12-\endcsname{12}
\expandafter\def\csname storetlsonethreeusetablebig-4-12-\endcsname{0.06\%}
\expandafter\def\csname storetlsonethreeusetablebig-5-12-\endcsname{10}
\expandafter\def\csname storetlsonethreeusetablebig-6-12-\endcsname{0.01\%}
\expandafter\def\csname storetlsonethreeusetablebig-7-12-\endcsname{12}
\expandafter\def\csname storetlsonethreeusetablebig-8-12-\endcsname{0.37\%}
\expandafter\def\csname storetlsonethreeusetablebig-9-12-\endcsname{9}
\def\tlsonethreeusetablebiglines#1#2{\ifcsname storetlsonethreeusetablebig-#1-#2-\endcsname\csname storetlsonethreeusetablebig-#1-#2-\endcsname\else\textbf{XxX}\fi}

\newcommand{\SYDALLCONNSexact}[1]{379,729,661\xspace}
\newcommand{\SYDALLCONNS}[1]{379.7M\xspace}
\newcommand{\SYDHTTPSCONNSexact}[1]{369,652,052\xspace}
\newcommand{\SYDHTTPSCONNS}[1]{369.7M\xspace}
\newcommand{\SYDVERSIONPREPARATIONexact}[1]{0\xspace}
\newcommand{\SYDVERSIONPREPARATION}[1]{0\xspace}
\newcommand{\SYDCIPHERPREPARATIONexact}[1]{0\xspace}
\newcommand{\SYDCIPHERPREPARATION}[1]{0\xspace}

\def\SYDnegotiatedversionstablelines#1#2{\ifcsname storeSYDnegotiatedversionstable-#1-#2-\endcsname\csname storeSYDnegotiatedversionstable-#1-#2-\endcsname\else\textbf{XxX}\fi}
\newcommand{\SYDnegotiatedversionscategory}[1]{%
\ifnum\pdfstrcmp{#1}{TLSv10}=0 1.6\xspace\else\ifnum\pdfstrcmp{#1}{TLSv11}=0 0\xspace\else\ifnum\pdfstrcmp{#1}{TLSv12}=0 88.7\xspace\else\ifnum\pdfstrcmp{#1}{TLSv13}=0 9.6\xspace\else %
\textbf{XXX}\xspace %
\fi\fi\fi\fi%
}%
\newcommand{\SYDsignaturetablepreparationexact}[1]{0\xspace}
\newcommand{\SYDsignaturetablepreparation}[1]{0\xspace}

\expandafter\def\csname storeSYDclientserverversionstable-1-1-\endcsname{TLSv10}
\expandafter\def\csname storeSYDclientserverversionstable-2-1-\endcsname{1.62\%}
\expandafter\def\csname storeSYDclientserverversionstable-3-1-\endcsname{35.72\%}
\expandafter\def\csname storeSYDclientserverversionstable-1-2-\endcsname{TLSv11}
\expandafter\def\csname storeSYDclientserverversionstable-2-2-\endcsname{0.04\%}
\expandafter\def\csname storeSYDclientserverversionstable-3-2-\endcsname{34.31\%}
\expandafter\def\csname storeSYDclientserverversionstable-1-3-\endcsname{TLSv12}
\expandafter\def\csname storeSYDclientserverversionstable-2-3-\endcsname{88.71\%}
\expandafter\def\csname storeSYDclientserverversionstable-3-3-\endcsname{96.47\%}
\expandafter\def\csname storeSYDclientserverversionstable-1-4-\endcsname{TLSv13}
\expandafter\def\csname storeSYDclientserverversionstable-2-4-\endcsname{7.64\%}
\expandafter\def\csname storeSYDclientserverversionstable-3-4-\endcsname{36.28\%}
\expandafter\def\csname storeSYDclientserverversionstable-1-5-\endcsname{TLSv13-7E01}
\expandafter\def\csname storeSYDclientserverversionstable-2-5-\endcsname{none}
\expandafter\def\csname storeSYDclientserverversionstable-3-5-\endcsname{0\%}
\expandafter\def\csname storeSYDclientserverversionstable-1-6-\endcsname{TLSv13-7E02}
\expandafter\def\csname storeSYDclientserverversionstable-2-6-\endcsname{none}
\expandafter\def\csname storeSYDclientserverversionstable-3-6-\endcsname{0\%}
\expandafter\def\csname storeSYDclientserverversionstable-1-7-\endcsname{TLSv13-draft18}
\expandafter\def\csname storeSYDclientserverversionstable-2-7-\endcsname{none}
\expandafter\def\csname storeSYDclientserverversionstable-3-7-\endcsname{0.04\%}
\expandafter\def\csname storeSYDclientserverversionstable-1-8-\endcsname{TLSv13-draft22}
\expandafter\def\csname storeSYDclientserverversionstable-2-8-\endcsname{none}
\expandafter\def\csname storeSYDclientserverversionstable-3-8-\endcsname{0\%}
\expandafter\def\csname storeSYDclientserverversionstable-1-9-\endcsname{TLSv13-draft23}
\expandafter\def\csname storeSYDclientserverversionstable-2-9-\endcsname{0\%}
\expandafter\def\csname storeSYDclientserverversionstable-3-9-\endcsname{0.15\%}
\expandafter\def\csname storeSYDclientserverversionstable-1-10-\endcsname{TLSv13-draft26}
\expandafter\def\csname storeSYDclientserverversionstable-2-10-\endcsname{0\%}
\expandafter\def\csname storeSYDclientserverversionstable-3-10-\endcsname{0.01\%}
\expandafter\def\csname storeSYDclientserverversionstable-1-11-\endcsname{TLSv13-draft27}
\expandafter\def\csname storeSYDclientserverversionstable-2-11-\endcsname{none}
\expandafter\def\csname storeSYDclientserverversionstable-3-11-\endcsname{0\%}
\expandafter\def\csname storeSYDclientserverversionstable-1-12-\endcsname{TLSv13-draft28}
\expandafter\def\csname storeSYDclientserverversionstable-2-12-\endcsname{0\%}
\expandafter\def\csname storeSYDclientserverversionstable-3-12-\endcsname{0.05\%}
\expandafter\def\csname storeSYDclientserverversionstable-1-13-\endcsname{TLSv13-FB23}
\expandafter\def\csname storeSYDclientserverversionstable-2-13-\endcsname{none}
\expandafter\def\csname storeSYDclientserverversionstable-3-13-\endcsname{0.01\%}
\expandafter\def\csname storeSYDclientserverversionstable-1-14-\endcsname{TLSv13-FB26}
\expandafter\def\csname storeSYDclientserverversionstable-2-14-\endcsname{1.98\%}
\expandafter\def\csname storeSYDclientserverversionstable-3-14-\endcsname{1.98\%}
\def\SYDclientserverversionstablelines#1#2{\ifcsname storeSYDclientserverversionstable-#1-#2-\endcsname\csname storeSYDclientserverversionstable-#1-#2-\endcsname\else\textbf{XxX}\fi}

\expandafter\def\csname storeSYDtlsonethreeusetablebig-1-1-\endcsname{google}
\expandafter\def\csname storeSYDtlsonethreeusetablebig-2-1-\endcsname{52.46\%}
\expandafter\def\csname storeSYDtlsonethreeusetablebig-3-1-\endcsname{1}
\expandafter\def\csname storeSYDtlsonethreeusetablebig-4-1-\endcsname{6.75\%}
\expandafter\def\csname storeSYDtlsonethreeusetablebig-5-1-\endcsname{4}
\expandafter\def\csname storeSYDtlsonethreeusetablebig-6-1-\endcsname{5.33\%}
\expandafter\def\csname storeSYDtlsonethreeusetablebig-7-1-\endcsname{3}
\expandafter\def\csname storeSYDtlsonethreeusetablebig-8-1-\endcsname{2.52\%}
\expandafter\def\csname storeSYDtlsonethreeusetablebig-9-1-\endcsname{5}
\expandafter\def\csname storeSYDtlsonethreeusetablebig-1-2-\endcsname{facebook}
\expandafter\def\csname storeSYDtlsonethreeusetablebig-2-2-\endcsname{24.72\%}
\expandafter\def\csname storeSYDtlsonethreeusetablebig-3-2-\endcsname{2}
\expandafter\def\csname storeSYDtlsonethreeusetablebig-4-2-\endcsname{1.54\%}
\expandafter\def\csname storeSYDtlsonethreeusetablebig-5-2-\endcsname{6}
\expandafter\def\csname storeSYDtlsonethreeusetablebig-6-2-\endcsname{1.15\%}
\expandafter\def\csname storeSYDtlsonethreeusetablebig-7-2-\endcsname{5}
\expandafter\def\csname storeSYDtlsonethreeusetablebig-8-2-\endcsname{0.12\%}
\expandafter\def\csname storeSYDtlsonethreeusetablebig-9-2-\endcsname{11}
\expandafter\def\csname storeSYDtlsonethreeusetablebig-1-3-\endcsname{other}
\expandafter\def\csname storeSYDtlsonethreeusetablebig-2-3-\endcsname{16.56\%}
\expandafter\def\csname storeSYDtlsonethreeusetablebig-3-3-\endcsname{3}
\expandafter\def\csname storeSYDtlsonethreeusetablebig-4-3-\endcsname{38.58\%}
\expandafter\def\csname storeSYDtlsonethreeusetablebig-5-3-\endcsname{1}
\expandafter\def\csname storeSYDtlsonethreeusetablebig-6-3-\endcsname{25.74\%}
\expandafter\def\csname storeSYDtlsonethreeusetablebig-7-3-\endcsname{2}
\expandafter\def\csname storeSYDtlsonethreeusetablebig-8-3-\endcsname{40.78\%}
\expandafter\def\csname storeSYDtlsonethreeusetablebig-9-3-\endcsname{2}
\expandafter\def\csname storeSYDtlsonethreeusetablebig-1-4-\endcsname{cloudflare}
\expandafter\def\csname storeSYDtlsonethreeusetablebig-2-4-\endcsname{5.79\%}
\expandafter\def\csname storeSYDtlsonethreeusetablebig-3-4-\endcsname{4}
\expandafter\def\csname storeSYDtlsonethreeusetablebig-4-4-\endcsname{1.04\%}
\expandafter\def\csname storeSYDtlsonethreeusetablebig-5-4-\endcsname{7}
\expandafter\def\csname storeSYDtlsonethreeusetablebig-6-4-\endcsname{64.17\%}
\expandafter\def\csname storeSYDtlsonethreeusetablebig-7-4-\endcsname{1}
\expandafter\def\csname storeSYDtlsonethreeusetablebig-8-4-\endcsname{4.44\%}
\expandafter\def\csname storeSYDtlsonethreeusetablebig-9-4-\endcsname{3}
\expandafter\def\csname storeSYDtlsonethreeusetablebig-1-5-\endcsname{amazon}
\expandafter\def\csname storeSYDtlsonethreeusetablebig-2-5-\endcsname{0.25\%}
\expandafter\def\csname storeSYDtlsonethreeusetablebig-3-5-\endcsname{5}
\expandafter\def\csname storeSYDtlsonethreeusetablebig-4-5-\endcsname{31.23\%}
\expandafter\def\csname storeSYDtlsonethreeusetablebig-5-5-\endcsname{2}
\expandafter\def\csname storeSYDtlsonethreeusetablebig-6-5-\endcsname{2.2\%}
\expandafter\def\csname storeSYDtlsonethreeusetablebig-7-5-\endcsname{4}
\expandafter\def\csname storeSYDtlsonethreeusetablebig-8-5-\endcsname{44.18\%}
\expandafter\def\csname storeSYDtlsonethreeusetablebig-9-5-\endcsname{1}
\expandafter\def\csname storeSYDtlsonethreeusetablebig-1-6-\endcsname{akamai}
\expandafter\def\csname storeSYDtlsonethreeusetablebig-2-6-\endcsname{0.15\%}
\expandafter\def\csname storeSYDtlsonethreeusetablebig-3-6-\endcsname{6}
\expandafter\def\csname storeSYDtlsonethreeusetablebig-4-6-\endcsname{5.29\%}
\expandafter\def\csname storeSYDtlsonethreeusetablebig-5-6-\endcsname{5}
\expandafter\def\csname storeSYDtlsonethreeusetablebig-6-6-\endcsname{0.34\%}
\expandafter\def\csname storeSYDtlsonethreeusetablebig-7-6-\endcsname{8}
\expandafter\def\csname storeSYDtlsonethreeusetablebig-8-6-\endcsname{4\%}
\expandafter\def\csname storeSYDtlsonethreeusetablebig-9-6-\endcsname{4}
\expandafter\def\csname storeSYDtlsonethreeusetablebig-1-7-\endcsname{squarespace}
\expandafter\def\csname storeSYDtlsonethreeusetablebig-2-7-\endcsname{0.03\%}
\expandafter\def\csname storeSYDtlsonethreeusetablebig-3-7-\endcsname{7}
\expandafter\def\csname storeSYDtlsonethreeusetablebig-4-7-\endcsname{0\%}
\expandafter\def\csname storeSYDtlsonethreeusetablebig-5-7-\endcsname{12}
\expandafter\def\csname storeSYDtlsonethreeusetablebig-6-7-\endcsname{0.03\%}
\expandafter\def\csname storeSYDtlsonethreeusetablebig-7-7-\endcsname{11}
\expandafter\def\csname storeSYDtlsonethreeusetablebig-8-7-\endcsname{0\%}
\expandafter\def\csname storeSYDtlsonethreeusetablebig-9-7-\endcsname{12}
\expandafter\def\csname storeSYDtlsonethreeusetablebig-1-8-\endcsname{ovh}
\expandafter\def\csname storeSYDtlsonethreeusetablebig-2-8-\endcsname{0.02\%}
\expandafter\def\csname storeSYDtlsonethreeusetablebig-3-8-\endcsname{8}
\expandafter\def\csname storeSYDtlsonethreeusetablebig-4-8-\endcsname{0.18\%}
\expandafter\def\csname storeSYDtlsonethreeusetablebig-5-8-\endcsname{9}
\expandafter\def\csname storeSYDtlsonethreeusetablebig-6-8-\endcsname{0.4\%}
\expandafter\def\csname storeSYDtlsonethreeusetablebig-7-8-\endcsname{7}
\expandafter\def\csname storeSYDtlsonethreeusetablebig-8-8-\endcsname{0.94\%}
\expandafter\def\csname storeSYDtlsonethreeusetablebig-9-8-\endcsname{8}
\expandafter\def\csname storeSYDtlsonethreeusetablebig-1-9-\endcsname{digitalocean}
\expandafter\def\csname storeSYDtlsonethreeusetablebig-2-9-\endcsname{0.01\%}
\expandafter\def\csname storeSYDtlsonethreeusetablebig-3-9-\endcsname{9}
\expandafter\def\csname storeSYDtlsonethreeusetablebig-4-9-\endcsname{0.11\%}
\expandafter\def\csname storeSYDtlsonethreeusetablebig-5-9-\endcsname{10}
\expandafter\def\csname storeSYDtlsonethreeusetablebig-6-9-\endcsname{0.5\%}
\expandafter\def\csname storeSYDtlsonethreeusetablebig-7-9-\endcsname{6}
\expandafter\def\csname storeSYDtlsonethreeusetablebig-8-9-\endcsname{1.16\%}
\expandafter\def\csname storeSYDtlsonethreeusetablebig-9-9-\endcsname{7}
\expandafter\def\csname storeSYDtlsonethreeusetablebig-1-10-\endcsname{azure}
\expandafter\def\csname storeSYDtlsonethreeusetablebig-2-10-\endcsname{0.01\%}
\expandafter\def\csname storeSYDtlsonethreeusetablebig-3-10-\endcsname{10}
\expandafter\def\csname storeSYDtlsonethreeusetablebig-4-10-\endcsname{15.07\%}
\expandafter\def\csname storeSYDtlsonethreeusetablebig-5-10-\endcsname{3}
\expandafter\def\csname storeSYDtlsonethreeusetablebig-6-10-\endcsname{0.05\%}
\expandafter\def\csname storeSYDtlsonethreeusetablebig-7-10-\endcsname{10}
\expandafter\def\csname storeSYDtlsonethreeusetablebig-8-10-\endcsname{1.21\%}
\expandafter\def\csname storeSYDtlsonethreeusetablebig-9-10-\endcsname{6}
\expandafter\def\csname storeSYDtlsonethreeusetablebig-1-11-\endcsname{godaddy}
\expandafter\def\csname storeSYDtlsonethreeusetablebig-2-11-\endcsname{0\%}
\expandafter\def\csname storeSYDtlsonethreeusetablebig-3-11-\endcsname{11}
\expandafter\def\csname storeSYDtlsonethreeusetablebig-4-11-\endcsname{0.01\%}
\expandafter\def\csname storeSYDtlsonethreeusetablebig-5-11-\endcsname{11}
\expandafter\def\csname storeSYDtlsonethreeusetablebig-6-11-\endcsname{0.02\%}
\expandafter\def\csname storeSYDtlsonethreeusetablebig-7-11-\endcsname{12}
\expandafter\def\csname storeSYDtlsonethreeusetablebig-8-11-\endcsname{0.45\%}
\expandafter\def\csname storeSYDtlsonethreeusetablebig-9-11-\endcsname{9}
\expandafter\def\csname storeSYDtlsonethreeusetablebig-1-12-\endcsname{alibaba}
\expandafter\def\csname storeSYDtlsonethreeusetablebig-2-12-\endcsname{0\%}
\expandafter\def\csname storeSYDtlsonethreeusetablebig-3-12-\endcsname{12}
\expandafter\def\csname storeSYDtlsonethreeusetablebig-4-12-\endcsname{0.21\%}
\expandafter\def\csname storeSYDtlsonethreeusetablebig-5-12-\endcsname{8}
\expandafter\def\csname storeSYDtlsonethreeusetablebig-6-12-\endcsname{0.08\%}
\expandafter\def\csname storeSYDtlsonethreeusetablebig-7-12-\endcsname{9}
\expandafter\def\csname storeSYDtlsonethreeusetablebig-8-12-\endcsname{0.19\%}
\expandafter\def\csname storeSYDtlsonethreeusetablebig-9-12-\endcsname{10}
\def\SYDtlsonethreeusetablebiglines#1#2{\ifcsname storeSYDtlsonethreeusetablebig-#1-#2-\endcsname\csname storeSYDtlsonethreeusetablebig-#1-#2-\endcsname\else\textbf{XxX}\fi}

\begin{abstract}

    TLS~1.3 marks a significant departure from previous versions of the
    Transport Layer Security protocol (TLS). The new version offers a
    simplified protocol flow, more secure cryptographic primitives, and new
    features to improve performance, among other things. In this paper, we
    conduct the first study of TLS~1.3 deployment and use since its
    standardization by the IETF. We use active scans to measure deployment
    across more than 275M domains, including nearly 90M country-code top-level
    domains. We establish and investigate the critical contribution that
    hosting services and CDNs make to the fast, initial uptake of the protocol.
    We use passive monitoring at two positions on the globe to determine the
    degree to which users profit from the new protocol and establish the usage
    of its new features. Finally, we exploit data from a widely deployed
    measurement app in the Android ecosystem to analyze the use of TLS~1.3 in
    mobile networks and in mobile browsers. Our study shows that TLS~1.3 enjoys
    enormous support even in its early days, unprecedented for any TLS version.
    However, this is strongly related to very few global players pushing it
    into the market and sustaining its growth.

\end{abstract}
\begin{document}
\maketitle
\section{Introduction}

The Transport Layer Protocol (TLS) is the backbone of secure communication over
the Internet. It is used to secure HTTP and, by extension, protocols operating
on top of HTTP. Applications running on servers, desktop devices, and mobile
phones all rely on it. Over the years the security of TLS has come
under increasing scrutiny and a long list of
vulnerabilities and flaws have been addressed in the last
decade~\cite{kotzias2018coming}. TLS~1.3,
the newest version of the protocol, redesigns central aspects, simplifying the
protocol, especially the handshake, and streamlining encryption to address many of these issues.
It also supports a higher degree of privacy by encrypting as early as possible
and improves performance by shortening the handshake.

Development of TLS~1.3 has been driven by major Internet corporations and
organizations, in particular Google, Facebook, and Mozilla, who felt a need to
address the security needs of their users and make their Web businesses faster
to access across all devices. This drive is in line with previous
contributions to TLS like Certificate Transparency, the HSTS
header for HTTP, and the downgrade protection SCSV. Previous work has found
evidence that the control that corporations like Google and Facebook
exercise---they control both endpoints of a connection---leads to new security
mechanisms being deployed faster~\cite{amann2017mission}.

% Johanna does not believe the following. This is not a problem.
% TLS~1.3 is different from the aforementioned
% technologies, however: while large corporations will find it easy to use their
% control over applications and server farms to deploy the protocol for their
% needs, smaller operators are at a disadvantage: when they need to make a
% decision whether and when to roll out TLS~1.3, they must carefully ascertain
% that they do not lose customers with older browsers. In sociological terms, TLS
% 1.3 has a high potential for market concentration, despite the many advantages
% it brings.

In this paper, we analyze both the use and deployment of TLS~1.3 employing
three different data sources: data from large-scale Internet scans,
passive traffic observation in the Northern and Southern hemisphere, and data
raised by a widely deployed application for the Android OS that can analyze TLS
handshakes. Our primary contributions are as follows:

\textbf{Deployment across DNS zones} We carry out large-scale scans of domains
across a large number of DNS zones, including \cno, 54 country-code top-level
domains (ccTLDs), and more than 1100 of the new generic TLDs allocated by
ICANN. To the best of our knowledge, this is the first time that TLS deployment
is analyzed with respect to the latter two categories. We show that deployment
varies significantly across the groups of domains we analyze.

\textbf{Observation of TLS~1.3 use} Our passive monitoring allows us to
identify the properties of TLS~1.3 traffic in more detail, including use of the
performance-enhancing improvements. We show the high degree of centralization
that this protocol represents as most connections terminate at hosts belonging to very few
entities.

\textbf{Impact of hosting} We enrich our data from both active and passive
measurement with DNS lookups and correlation with IP ranges of large cloud
providers known to host a significant amount of domains. We show that the
domain front-end service Cloudflare is dominant in the case of deployed TLS
1.3; however, in TLS~1.3 use we find Facebook and Google to be responsible for
a majority of connections.

\textbf{The Android ecosystem} Finally, we use data from the Lumen
privacy-enhancement app to investigate under which circumstances mobile devices
make use of TLS~1.3. We discover that large corporations have experimented with
various, different versions of TLS~1.3, which are sometimes still in use but
not identical to the standardized version.

\section{Background}

TLS~1.3 was standardized by the IETF in August 2018~\cite{rfc8446}. It presents
a departure from previous versions of the protocol in order to avoid inheriting
the flaws and vulnerabilities present in older versions
~\cite{kotzias2018coming,logjam,lucky13,197245,moller2014poodle}.  TLS~1.3
brings two main advantages: enhanced security and improved speed. The former is
achieved thanks to a new protocol flow and modern cryptographic algorithms,
including mandated Perfect Forward Secrecy and allowing only symmetric ciphers
that provide authenticated encryption (AEAD ciphers). 
% The old RC4 stream cipher, which was broken in 2013, has a successor in Bernstein's ChaCha20. 
The speed also improves due to changes in the handshake that reduce the number of
necessary round-trips. Encrypted payloads are now commonly sent after 
just one round-trip (1-RTT). In some cases where a past cryptographic key can be
re-used, data can be sent in the first TCP packet (0-RTT mode)~\cite{tls13-0rttCloudflare}.
TLS~1.3 also improves user privacy and encrypts the payload as early as possible. This
includes the server certificate and many extensions (especially on the server side)
which were sent in the clear in the past.

As opposed to previous TLS versions, stakeholders began supporting and
experimenting with TLS~1.3 variants very early, long before the standardization
by the IETF had finished. This early adoption was driven by big actors like Mozilla,
Cloudflare, Google, and---at lower intensity---Facebook.  Cloudflare became the
first cloud and CDN provider enabling TLS~1.3 for its customers as early as
September 2016~\cite{tls13}. Between December 2017 and May 2018, the
percentage of TLS~1.3 connections served by Cloudflare grew from 0.06\% to
5-6\%~\cite{yougettls13everyonedoes}. Facebook has
deployed TLS~1.3 support globally in their apps and clients (including WhatsApp and
Instagram) and in their user-facing and internal infrastructure. They report
more than 50\% of their global Internet traffic being TLS~1.3 as of August
2018~\cite{fbFizzDeployment}.
Other actors have not yet made clear announcements: Akamai, for
instance, announced support for TLS~1.3 to start in mid-2018~\cite{akamaiTLS},
but to our knowledge has not yet enabled TLS~1.3 support.

Many end-user clients are already TLS~1.3-enabled. Mozilla's Firefox and
Google's Chrome were the first browsers to support TLS~1.3 in March 2017
(Firefox v52) and December 2017 (Chrome v56), respectively. Both initiated
support in February 2017 with a beta roll-out for a fraction of their customers.
Due to incompatibilities with popular middleboxes such as BlueCoat
proxies~\cite{bluecoatTLS13}, discovered in this unprecedented trial program,
they deprecated support for a time until this was resolved.

% \todo{I am not sure we should keep the long history in, apart from the
% middlebox fiasco, which motivated \textit{supported versions} Candidate for
% shortening.} They resumed interrupted and full suport with the release of
% versions \missingno and v66 (April 2018), respectively. In August 2018,
% Mozilla's telemetry indicated that around 5\% of Firefox connections were TLS
% 1.3~\cite{mozillaTLS}. There is wide support for TLS~1.3 in open source
% libraries, although this will only become helpful once major distributions of
% open source OSes ship binaries that link to these new versions. wolfSSL
% supported TLS~1.3 since 3.11.1 (May 2017), OpenSSL since v1.1.1 (September
% 2018). Facebook has open-sourced its TLS~1.3 C++ library Fizz~\cite{fbFizz},
% which is a part of their open-source HTTP framework Proxygen~\cite{fbProxygen}.
% Mozilla's Network Security Services (NSS) have supported TLS~1.3 since February
% 2017~\cite{mozillaNSS}.

% \todo{Lumen talks more about Client Hello (CH) and SH. We might need to
% introduce htem here. I think passmon might also profit.
% Johanna: passmon doesn't use that distinction much; might try to kill from Lumen}
% 

\section{Related Work}

Many academic studies have characterized and studied different aspects of the
TLS and X.509 PKI ecosystem, including the general state of the ecosystem and certificate
validation~\cite{clark2013sok, sslerrors,
noattacknecessary,durumeric2013zmap, holz11imc}, certificate
revocation~\cite{amann2017mission,debiankeys, wake_of_heartbleed},
vulnerability discovery~\cite{logjam,lucky13,197245,moller2014poodle},
Certificate Transparency~\cite{scheitle2018rise,vandersloot2016towards,
ctgossip, ctemail, firstlookct}, and TLS/HTTPS
support~\cite{holz2015tls,scheitle2018measuring}. In the case of TLS~1.3
many studies examined the protocol and proposed improvements and new features
~\cite{krawczyk2016optls,badertscher2015augmented,
bhargavan2017implementing,delignat2017implementing}, cryptographic
schemes~\cite{bellare2016multi}, performed protocol verification and cryptographic
analysis~\cite{dowling2015cryptographic,kobeissi2018formal,
bhargavan2017verified,beurdouche2015flextls} (including symbolic
analysis~\cite{cremers2017comprehensive}) to discover vulnerabilities and
flaws~\cite{jager2015security}. 

More closely related to our work, only a limited number of previous
studies measure TLS~1.3 deployment and support. However,
none of these papers focus on performing a comprehensive analysis
of TLS~1.3. Instead, they gain anecdotal insights about ongoing
deployment efforts of TLS~1.3 as a by-product of their attempts to answer more general research
questions about TLS deployment in the wild. Kotzias et
al.~\cite{kotzias2018coming} perform a longitudinal analysis of TLS
deployment for five years. In their work, the authors focus on changes
in TLS deployment caused by the disclosure of protocol vulnerabilities.
The authors also briefly report on TLS~1.3 deployment in April 2018 (23\% of client
connections supporting it while it being used in $\approx$ 1\% of connections). 
TLS~1.3 was not the focus of their study. Note that Kotzias \etal also use data from
the ICSI SSL Notary for their analysis.
%
% Johanna thinks this is boring.
% The authors suggest that this early support was
% mainly driven by newer versions of the popular Chrome and Firefox browsers
% being rolled out ~\footnote{Firefox version~60.0, released in May 2018
% ~\cite{firefox60releaseNotes}}  and the support of major CDN providers (\eg
% CloudFlare). 
A 2017 study analyzing Lumen data (which we also use in our paper) reported 
marginal support of TLS~1.3 extensions~\cite{razaghpanah2017studying}.
This early support was driven mainly by
Android software developed by large companies like Facebook. Finally, a number
of studies have focused on QUIC --an UDP-based protocol that has been proposed
as a lightweight and low-latency alternative to TLS-based
protocols~\cite{cui2017innovating,ruth2018first,langley2017quic,kakhki2017taking,lychev2015secure}.

\section{Datasets and Methodology}

In the following, we describe our data collection from three sources.
We choose our data sources to cover as
many angles of the burgeoning TLS~1.3 ecosystem as possible. Our primary data
sources are active scans to capture deployment, passive monitoring to
understand actual use in practice, and analysis of mobile phone traffic on the
device to understand the difference to the mobile world. We enrich our data
sets with lookups of IP ranges of important cloud providers and DNS scans to
obtain nameserver records. We use these to determine the importance of big
stakeholders for both deployment and use of TLS~1.3.
Please note that we discuss the ethical considerations of our data collections
in Appendix~\ref{sec:appendix_ethics}.

% XMETH marker
\subsection{Active scanning}
\label{sub:active-scanning}

We perform active scans of Internet domains to measure the deployment of TLS
1.3. We reuse and extend a method previously published in \cite{Amann2017}. 
Our scanning is performed from a large research university in Australia.

\noindent \textbf{Obtaining domain lists} On 1 May 2019, we collect a large
number of domain names from publicly available sources. Our input sets are
zone files plus three top lists, namely the Alexa Top 1M list, the Majestic list, and Cisco's
Umbrella list.  Scheitle \etal analyzed the composition of top
lists in \cite{Scheitle2018}; our choice of the Alexa list and
\textit{com/net/org} is based on their recommendations for lists with
mostly functional Web sites and lists representing a general
population. We obtain the zone files for \textit{com} and \textit{org}
directly from the operators, and \textit{net} from ICANN's Centralized Zone
Data Service, together with 1120 zone files of the new generic TLDs (gTLDs
hereafter). From ViewDNS\footnote{https://viewdns.info}, we acquire domain
lists for 54 country-code TLDs (ccTLDs). ViewDNS claims to base these on Web
crawls, updating them at least every few months. We accept the bias towards Web
domains and verify that we have a sufficiently high number of domains in each
ccTLD.  Only 9 ccTLDs have fewer than 100k domains, but none has less than 19k.
For 22 ccTLDs , we have more than 500k domains, and for 12 ccTLDs more than 1M
domains (including \tld{fr}, \tld{ru}, \tld{cn}, \tld{eu}, \tld{nl}, and
\tld{de}). We lament the absence of \tld{uk}.

We combine our input lists into three domain lists of increasing size, each for
one scanning campaign: one for the Alexa domains, one for the ccTLDs, and one
for all other domains. In the post-processing stage, we subdivide the results
for the latter into \textit{com/net/org} and the new gTLDs. Table
\ref{tab:input-lists} shows the number of domains in each input set. As some
names in the top lists include subdomains and not just the registrable name
under the TLD (\eg \textit{sub2.sub1.example.com.au}), we use Mozilla's public
suffix list to derive the registrable domain name (\textit{example.com.au})
and add it our input list.

% zcat 1556701596_com.domain.sortu.csv.gz | wc -l 140203507
% zcat 1556701596_net.domain.sortu.csv.gz | wc -l 13614440
% zcat 1556701596_org.domain.sortu.csv.gz | wc -l 10149832
% zcat 1556701596_cno.domain.sortu.csv.gz | wc -l 163967779

% zcat 1556701596_czdsnonet.domain.sortu.csv.gz | wc -l 23429945
% zcat 1556944522_cctld.domain.sortu+2ld.csv.gz | wc -l 87933097

\begin{table}[tb]
    \small
    \centering
    \caption{Domain lists in each scanning campaign.}
    \label{tab:input-lists}
    \begin{tabular}{lrr}
        \toprule
        Input data set              & \# Domains    & Campaign dates    \\
        \midrule
        Alexa                       & 1.0M          & 1 May 2019        \\
        ccTLDs                      & 87.93M        & 4-5 May 2019      \\
        \midrule
        \textit{com/net/org}        & 163.97M       & 1-3 May 2019      \\
        gTLDs                       & 23.43M        &                   \\
        \bottomrule
    \end{tabular}
\end{table}

\noindent \textbf{Resolving domains, port scanning} We resolve all domains to A
records using Scheitle's fork of the \textit{massdns}
tool\footnote{\url{https://github.com/quirins/massdns}}. We resolve CNAMEs up
to 15 levels of indirection. We then run \textit{zmap}~\cite{durumeric2013zmap}
to identify all IP addresses with open port TCP/443.

\noindent \textbf{TLS scans} We rebase and modify \textit{goscanner}, the TLS
scanner by the authors of \cite{Amann2017}, to carry out a TLS handshake for
every domain hosted on such an IP address, enabling support for TLS~1.3 and
sending this protocol version as first preference.  We create a full PCAP of
all scans using \textit{tcpdump}. We also use Zeek (formerly Bro) to
parse the PCAP files. This enables us to investigate failed handshakes in more
depth than previous work.

Basing our scans on domain names rather than IP addresses allows us to support
the Server Name Extension of TLS, \ie HTTP virtual hosts. It also allows us to
identify differences between domains in different groups (\eg gTLDs vs ccTLD vs
Alexa) and identify countries with particularly high deployment. Furthermore, it
enables identification of differences per domain, \eg when a domain configures TLS~1.3 on
one IP address, but not on others.

% Not sure what we want to mention here - also out of time.
% \todo{Say which extensions we do check?}

\noindent \textbf{Limitations} Our active scans do not investigate TLS deployment on IPv6:
our hosting institution in Australia does not yet offer general IPv6
connectivity. We also do not check support for many of the new extensions that TLS~1.3
defines, \eg Certificate Authorities or Post-Handshake Client Authentication,
leaving this to future work. 
% This is so common that Johanna does not think it is worth mentioning
%We note that our scanner is written in Go, and the current TLS library only supports a subset of extensions as well. Servers must not send extensions that the client did not request,
%either. Hence, this is a minor limitation for us.

\subsection{Passive observations}
\label{sec:passivemeth}

Our passive data collection uses two data sources. \first We have access to
data from the ICSI SSL Notary~\cite{notarypaper}, a large-scale observation
effort of TLS that began in 2012, monitoring at sites mostly located in
Northern America. \second To enrich our analysis with additional data, we also
collect data from the campus of a large Australian research university with
more than 50,000 students. Both the ICSI Notary as well as our Australian data
collection efforts use the Zeek Network Security Monitor~\cite{zeek} (until
recently known as Bro) to collect their data.

Since its inception in February 2012, the Notary has observed more than
400 billion TLS connections; a number of institutions has contributed, typically
5--8 different sites contribute data simultaneously. The data collected has been expanded over the years,
as the TLS protocol changed. It is, however, difficult to quickly adapt this
data collection effort for new protocol features. There are several reasons.
This data collection is run in operational environments using Zeek to secure
their networks. The collection effort is hence a best-effort service by the
operators and can only use the data that is provided by the current Zeek
version used on-site. It is thus not possible to quickly collect data on new
features or to change the data collection to answer emerging research
questions. Furthermore, expanding the data that the Notary collects typically
needs the data collection to be re-approved.

Thus, we amend our data collection effort by collecting four full days worth
of data (2019-05-09 16:00 till midnight 2019-05-13, local time) at a large research university in Australia. This data collection
effort is directly under our control and collects additional information from
the handshake, like the presence of TLS~1.3 Hello Retry requests (see Section~\ref{sec:earlydata}).
Furthermore, we use it for a geographic comparison of the
TLS~1.3 traffic. During the 4 days of data
collection, we saw \SYDALLCONNS{} TLS connections.  Note that while the Notary
dataset contains IPv4 and v6 traffic, the Australian University does not yet
support IPv6; we thus do not encounter IPv6 traffic in this measurement.
%\footnote{This collection effort was originally meant to last for a week;
%however, one of the authors of the paper accidentally irrecoverably deleted
%several days of data.}

\textit{Limitations} We note that for the ICSI Notary, the dataset exhibits
artefacts of the collection process that are beyond our control. As the Notary
leverages operational setups that run the analysis on top of their normal
duties, one must accept occasional outages, packets drops (\eg due to CPU
overload) and occasional misconfigurations. As such, the Notary data collection
effort is designed as a ``best effort'' process: it aims at as much coverage as
possible, but we can usually not quantify what it misses. Given the large total
volume across our sites, however, we consider the aggregate as
representative of many properties of real-world TLS activity.

While the Notary collected data from outside North America in the past, currently
all contributing sites are inside of North America. Furthermore, only one large
University Campus (with more than 30,000 students) is currently contributing the
full set of TLS~1.3 data items the Notary can collect. Some of the Notary analysis
thus only uses data from this single site---we make this explicit 
by referring to the site as $N1$ when this is the case.

\subsection{Lumen}
\label{sec:lumendescription}

The Lumen Privacy Monitor~\cite{razaghpanah2017studying} is a privacy-enhancing tool for Android, available on Google
Play~\cite{razaghpanah2015haystack}. Lumen intercepts and analyzes mobile traffic in user space (and
on localhost) to help users stay on top of their mobile traffic and privacy by
reporting network flows and personal data dissemination and allowing them to
block undesired traffic. Making Lumen available to the public as a
privacy-enhancing solution allowed the project to recruit users from all over the world,
and as a result to collect a large amount of anonymized real-world traffic
data generated by real user stimuli.

\noindent \textbf{Mobile traffic interception in user-space} Lumen acts as
middleware between apps and the network interface: it leverages the Android VPN
permission and implements a complete but simplified network stack to capture
and analyze network traffic locally without requiring root permissions. Lumen
collects and inspects mobile traffic transparently, regardless of the transport
and application-layer protocol used by a mobile application without
modifying the network path. Lumen is able to correlate traffic flows with app
identifiers and process IDs: it can accurately match TLS flows to the
process that generated them.

\noindent \textbf{Dataset} Between November 2015 and April 2019, more than
22,000 users from over 100 countries installed Lumen from Google Play.
Lumen's dataset contains accurate yet anonymized traffic fingerprints for more
than 92,000 Android apps, excluding mobile browsers to preserve users'
anonymity (see the ethical discussion in Appendix~\ref{sec:appendix_ethics}). For
this paper, we analyze 11.8 million TLS connections (Client and Server Hello
records) from 56,221 apps connecting to 149,389 domains (identified by the
Server Name Indication field).
% Removing ports to not get reviewers to ask questions about this :)
% across 417 TCP ports. 
%
% Also removing this, because anything else does not make sense.
%To minimize bias, we
%only study the Client Hello records before interception by Lumen's transparent
%TLS interception proxy, as generated by the app, and the Server Hello records
%directly received from the server. 

The $>$ 92,000 apps in Lumen's dataset include
many different types: apps downloaded from Google Play (2\% of the apps have
more than 1M installations according to Google Play's metadata), pre-installed
software~\cite{gamba2020analysis}, and apps downloaded from alternative app stores (\eg
F-Droid). Furthermore, Lumen collects data from many different OS versions.

% Johanna: too long and doesn't fit here.
% crowdsourcing data collection strategy also
% allows to study mobile apps across versions and different OS versions. The
% latter feature is critical as most app developers use the TLS libraries offered
% by default by the Android platform, which changes significantly across versions
% (not only version support but also preferred cipher suites, TLS extensions,
% \etc).  In August 2017, a study of over 5,000 Android apps revealed that about
% 84\% of them use default OS APIs for TLS~\cite{razaghpanah2017studying}.

\subsection{Hosting and CDNs}
\label{sub:hosting}

As the development of TLS~1.3 is very much driven by industry players offering
different Web services, a key element of our study is the impact that cloud and
hosting providers have on the deployment of TLS~1.3. 
We are not aware of a curated list containing the IP ranges of major
providers and what they are used for.  We use the term `cloud hosting' relatively
imprecisely, acknowledging that there are many different, often overlapping
forms---from securing front-ends (\eg Cloudflare) and `classic' provisioning of
an entire Web presence (Square\-space, GoDaddy), to CDNs. In this paper, we group
these different providers together when they share one property: they are in
control of a public server TLS endpoint intended for Web users, and hence they
can control which TLS version is used.

We searched for IPv4 and IPv6 blocks on the websites of arguably the most
common cloud providers. Cloudflare, Amazon AWS, and Microsoft Azure disclose
the IP allocations for their services. DigitalOcean, Google, and Alibaba do
not. We obtain their IP ranges via the search interface of Hurricane Electric's
\textit{bgp.he.net}. To minimize the false positive rate, we manually exclude
IP blocks that, from the description, either belong to cooperation partners
like ISPs, are access networks (like Google/Alphabet Fibre), or are intended
for corporate use or data caches and hence are unlikely to host a front-end to
a website.

Most, but not necessarily all domains in a given provider range should be
considered `hosted'. Similarly, a provider's IP complete ranges may not be
found via the mentioned search interface. Cloudflare's primary business models
are DNS provisioning and acting as a more secure front-end for Web services. A
(non-Cloudflare) domain with an IP address in Cloudflare's range is very likely
set up to use Cloudflare's Web front-end.  A similar argument holds for
Squarespace as a dedicated website creation platform: domains with IPs in this
range are very likely hosted. However,  DigitalOcean, Amazon AWS, Azure, and
Alibaba Cloud all offer products around `elastic' computing, allowing customers
to spin up virtual machines from provided images. They offer many other
products as well, however. Their IP ranges will almost certainly contain some
domains belonging to provider infrastructure, although their number should be
very small compared to the millions on our domain lists. Google and GoDaddy are
even more complex cases. Both have hundreds of IP ranges registered but also
own many subsidiaries, whose ranges may not be listed under the name of the
parent company. The number of their internal domain names should dwindle
compared to the size of our domain lists; however, we expect to miss out on
some hosted domains as we do not know the ranges of the subsidiaries.

We also add further providers of a different kind. We add the IP ranges for
Akamai for comparison. Due to Akamai's strategy of intelligently mirroring
customer content, our hypothesis is that most of these IPs will not serve as
front-end servers for Web sites, but it is worth validating this. We include a
VPS provider from Europe, OVH. The origin of the provider is not the `elastic'
model; it classically attracts customers who want a high degree of
customization.  Finally, we retrieve the allocated ranges for Facebook. We use
these to identify connections to Facebook services in our passive monitoring.

In addition to IP ranges, we employ a second method to identify hosting setups
that do not use one of the described providers. For every domain where a TLS
handshake in our active scan used TLS~1.3, we also retrieve the nameserver (NS)
record with \textit{massdns}.

% XRESULTS jump mark for vim
\section{Results}
\label{sec:results}

We present deployment findings using our active scans; we then turn to use of
TLS~1.3 using passive monitoring and Lumen.

\subsection{Deployment}
\label{sec:results:deployment}

\subsubsection{Incomplete handshakes} Since ca. 2011, reports such as
\cite{holz11imc,Durumeric2013,Amann2017} have consistently shown that servers
with open port TCP/443 often do not complete a TLS handshake. Depending on the
choice of scanning targets (domains or Internet-wide scans), this can be around
27-33\% of hosts. With domain data across many zones available, we investigate
this in more depth than the mentioned previous publications. For Alexa domains,
the rate we determine is only 4\%. For \cno, the gTLDs, and the ccTLDs, it is
17-19\%. The ccTLDs offer a markedly different picture: we find a high failure
rate in some well-known TLDs like \tld{cn}, \tld{de}, \tld{eu} (44\%, 39\%,
24\% respectively), the rate being between 5-15\% for roughly half the
ccTLDs. Roughly a quarter of ccTLDs show a failure rate between 1-5\%---
interestingly, we do not find `important' ccTLDs in there, with the possible
exceptions of \tld{pl} (Poland), \tld{au} (Australia), and \tld{arg}
(Argentina). The ccTLDs for which we have the fewest samples are part of the
middle group---we hence have no reason to believe that ViewDNS's collection
process introduced a bias.

While in many cases the TLS connections are aborted without any TLS protocol
messages by the server, in some cases the server sends a TLS alert before the
connection is established. For
example, 6.8\% of connections to ccTLDs, and 1.8\%, 2.9\%, and 2.3\% to \cno
domains abort connections with an \emph{unrecognized name} alert, signifying
that the server does not accept the domain name that we send it. Manually
contacting a small subset of these servers yields the same result when no SNI
is sent. An \emph{internal error} alert signifying a server problem is also
common, appearing in 1.2\% of ccTLD connections and 1.8\%, 2.2\%, 2.3\% of \cno
domains. The generic alert \emph{handshake error} appears in 0.5\%-1\% of
connections. Other alerts (like \emph{protocol version} or \emph{certificate
error}) are much less common, only appearing in a few hundred connections.

Our findings are not conclusive but point at different deployment strategies
utilizing different hosting providers in different countries. This is supported
by our later analysis of common hosters.

\begin{table*}[tb]
    \caption{Overview of TLS deployment across zones; percentages given in relation to resolvable domains, \ie those with an A record. Note that numbers do not add up to 100\% as we do not include failed handshakes, \eg due to server problems.}
    \label{tab:deployment-across-zones}
    \small
    \centering
    \begin{tabular}{lrrrr}
        \toprule
                                    &   \# Alexa Top 1m (\%)   &  \# \cno (\%)              & \# new gTLDs (\%)         & \# ccTLDs (\%) \\
        \midrule
        Resolved domains            &   940.5K (100\%)         &  144.0M (100\%)            & 17.5M  (100\%)            & 72.4M (100\%) \\
        \ldots open port 443        &   836.0K (88.89\%)       &  79.2M (55.01\%)           & 6.8M (39.07\%)            & 39.0M (53.90\%) \\
        \ldots with TLS~1.3         &   174.3K (18.53\%)       &  7.8M  (5.38\%)            & 1.3M (7.62\%)             & 3.3M  (4.54\%)  \\
        \ldots with TLS~1.2         &   613.0K (65.18\%        &  54.7M (38.01\%)           & 4.0M (22.99\%)            & 27.2M (37.64\%) \\
        \ldots with TLS 1.1         &   217 (0.02\%)           &  6.8K     (0.0\%)          & 153   (0.0\%)             & 11.4K    (0.02\%)  \\
        \ldots with TLS~1.0         &   17.5K (1.86\%)         &  2.0M  (1.38\%)            & 178.1K (1.02\%)           & 1.5M  (2.10\%)  \\
        \midrule
        IP addresses                &   584.8K (100\%)         &  10.8M (100\%)             & 1.5M  (100\%)             & 4.4M (100\%) \\ 
        \ldots open port 443        &   521.5K (89.17\%)       &  5.1M  (47.38)             & 779.0K (53.51\%)          & 2.6M (58.86\%)  \\
        \ldots with TLS~1.3         &   101.1K (17.29\%)       &  299.2K   (2.76\%)         & 100.3k (6.89\%)           & 196.6K (4.43\%) \\
        \ldots with TLS~1.2         &   394.3K (67.42\%)       &  3.9M  (36.4\%)            & 578.9K (39.77\%)          & 2.1M (46.87\%)\\
        \ldots with TLS 1.1         &   197 (0.03\%)           &  2.5K     (0.02\%)         & 113    (0.01\%)           & 1.3K (0.03\%)   \\
        \ldots with TLS~1.0         &   14.8K (2.53\%)         &  266.4K   (2.46\%)         & 16.7K  (1.15\%)           & 127.7K (2.88\%) \\
        \bottomrule
    \end{tabular}
\end{table*}

\subsubsection{Deployment across DNS zones}

We present our findings across our chosen domain groups. Table
\ref{tab:deployment-across-zones} summarizes TLS versions across servers in
different DNS zones and across the corresponding IP addresses. Note that the
percentages are given as percentages of \textit{resolvable domains}.

We test for deployment differences within our largest group, \cno, but find
that the respective percentages are never more than 1-2\% different between the
zones. As one would expect, Alexa domains almost always offer an open HTTPS
port, and roughly half of the domains in our ccTLDs and \cno do so, too. The
new gTLDs lag behind---however, many of these TLDs are known to be parked, or
have been acquired by large corporations to protect their DNS names, \ie they
are not intended for public, external access~\cite{Halvorson2015}.

We say a domain supports TLS~1.3 if at least one of its IP address supports
this version. TLS~1.3 is best supported on Alexa domains at 18.5\%. Only ca.
5\% of \cno and ccTLDs support it. Interestingly, this percentage is slightly
higher for the new gTLDs. The aforementioned, common use of gTLD domains could
be a reason, but also different hosting choices. \todo{RH: unfortunately, we
did not have the time to investigate this more deeply. Shorten?} Across all
zones, TLS~1.2 is much better supported, showing that the roll-out of the new
version is only picking up. Support for TLS 1.1 and 1.0 has fallen to
negligible levels. Note that we do not scan SSL 3, which is reported by our
scanner as a failed handshake. 
\subsubsection{Inconsistent use of protocol}

For Alexa domains, we verify how often domains that support TLS~1.3 do not
configure it for all IP addresses. We find only 105 Alexa domains where a
different TLS version is configured on an alternative IP, and in less than a
handful of cases the other protocol is TLS~1.0 and not TLS~1.2.  This
corresponds to 0.01\% of domains with successful handshakes. Handshakes failing
on the alternative IP address is more common (356 domains).  We also
investigate ccTLDs and the new gTLDs; we find the same percentage as for Alexa.
It is twice that high for \cno, but we conclude that overall domains deploy TLS
versions remarkably consistently.

\subsubsection{Server preferences for ciphers}

TLS~1.3 defines just five cipher suites. We offer the three supported by Go:
128-bit AES in GCM mode, ChaCha20 Poly1305, and 256-bit AES in GCM mode
in this order. 128-bit AES is used in the overwhelming amount of
connections: 90\% for domains in the Alexa list, in \cno, and in the new gTLDs
with most of the remaining connections using 256-bit AES. For ccTLDs 256-bit
AES was more popular (29.6\%) and ChaCha20 got a bit more than 2\%. This could
again point at different deployment strategies. ChaCha20 is very unpopular despite
being Google's choice for a secure stream cipher.

\subsubsection{TLS~1.3 by ccTLD}

We investigate the use of TLS~1.3 in the ccTLDs.
Table~\ref{tab:tls13-by-cctld} shows deployment of TLS~1.3 by ccTLD as a
percentage of TLS connections. The range is wide: at the top, we find TLDs with
75-80\% TLS1.3---they are \tld{cf} and \tld{tk}.  The East European countries
Ukraine, Slovakia, and Poland follow---but at much lower deployment (27-42\%).
On the next five ranks, we find Denmark, but also popular TLDs like \tld{io}
and \tld{me}. At the bottom end, surprisingly, we find large European zones and
economically strong countries: Germany, France, and Japan.

\begin{table*}
    \caption{Deployment of TLS~1.3 across 54 ccTLDs. Percentages indicate fraction of all TLS connections. Note that \tld{rf} is our transliteration for \tld{xn--1ai}, \ie the Russian Federation.}
    \label{tab:tls13-by-cctld}
    \small
    \centering
    \begin{tabular}{llr|llr|llr|llr|llr|llr}
        \toprule
         1 & \tgb{cf} & 80.1\%      & 11 & au & 17.7\%      & 21 & nz & 11.8\%      & 31 & ir & 9.0\%   & 41 & kz & 6.4\%   & 51 & \trb{de} & 3.8\% \\
         2 & \tgb{tk} & 75.0\%      & 12 & ma & 17.7\%      & 22 & ru & 11.1\%      & 32 & rf & 8.5\%   & 42 & cl & 6.3\%   & 52 & \trb{za} & 3.3\% \\
         3 & \tgb{ua}  & 42.3\%      & 13 & ro & 15.9\%      & 23 & sg & 11.1\%      & 33 & nl & 8.2\%   & 43 & mx & 5.7\%   & 53 & \trb{fr} & 3.2\% \\
         4 & \tgb{sk}  & 40.0\%      & 14 & co & 15.7\%      & 24 & ie & 10.7\%      & 34 & cz & 8.1\%   & 44 & rs & 5.4\%   & 54 & \trb{jp} & 2.8\% \\ 
         5 & \tgb{pl}  & 28.0\%      & 15 & la & 14.3\%      & 25 & at & 10.4\%      & 35 & pe & 7.5\%   & 45 & ar & 4.7\%   & & \\ 
         6 & dk       & 25.8\%      & 16 & il & 14.3\%      & 26 & eu & 10.2\%      & 36 & br & 7.4\%   & 46 & be & 4.7\%   & & \\ 
         7 & io       & 22.6\%      & 17 & cc & 13.0\%      & 27 & tv & 10.1\%      & 37 & in & 6.8\%   & 47 & no & 4.6\%   & & \\
         8 & me       & 22.4\%      & 18 & tr & 12.5\%      & 28 & my & 10.0\%      & 38 & es & 6.7\%   & 48 & se & 4.5\%   & & \\
         9 & us       & 21.8\%      & 19 & gr & 12.1\%      & 29 & ca & 9.5\%       & 39 & it & 6.6\%   & 49 & hu & 4.4\%   & & \\
         10 & cn      & 19.0\%      & 20 & su & 11.9\%      & 30 & ch & 9.4\%       & 40 & tw & 6.5\%   & 50 & \trb{pt} & 4.1\%   & & \\
        \bottomrule
    \end{tabular}
\end{table*}

We investigate the top 5 and bottom 5 ccTLDs more closely. Both \tld{cf}
(Central African Republic) and \tld{tk} (Tokelau) are well-known for allowing
the creation of domain names at no cost. The high numbers for these domains are
easy to explain. Domains in \tld{cf} resolve to 56.7k distinct IP addresses. Of
these, an impressive 49.6k are in IP ranges of the hosters we
identified---almost exclusively Cloudflare (48.0k). The situation is similar in
\tld{tk}: of the 72k IPs, 53k are in the ranges of our hosters.

Ukrainian domains resolve to 33.7k IP addresses; however, only 12.1k lie in our
hosting ranges. Cloudflare dominates(9.4k), followed by DigitalOcean ($\approx$
900), OVH ($\approx$ 800), and Amazon ($\approx$ 700). To better understand the
possible reasons for this high deployment, we inspect under which second-level
domain the DNS nameservers of \tld{ua} of domains that serve TLS~1.3 are
operating. We find that 19\% are operated by PromDNS, a GoDaddy subsidiary, and
51\% by Inhosted, a hosting company in Scotland. Together with Cloudflare,
these providers are responsible for the majority of TLS~1.3 in this TLD. We
find similar market concentration for \tld{sk}, where 78\% of nameservers
belong to the hosting company \textit{websupport.sk}, and \tld{pl}, where 67\%
belong to the domain hoster \textit{nazwa.pl}.

At the lower end of Table~\ref{tab:tls13-by-cctld}, we find a market
concentration for Cloudflare: 59\% of nameservers belong to the company in the
case of Portugal (\tld{pt}) and South Africa (\tld{za}), and 60\% in the case
of France.  In Germany, Cloudflare and 1blu dominate (30\% and 28\%,
respectively). Japan is different again: 56\% come from the Japanese hoster
\textit{value-domain.com}, and 19\% from Cloudflare.

\noindent \textit{Caveat.} It is important to note that our data does not
contain all name servers in the respective zones, but only those for domains
with TLS~1.3. As most domains in \tld{cf} and \tld{tk} are hosted by
Cloudflare, we can say with confidence that Cloudflare is also the reason for
the high deployment of TLS~1.3. However, in most other cases, our data does not
allow us to identify conclusive reasons for high or low TLS~1.3 deployment
beyond our chosen, large cloud providers.

%It is entirely possible that many domains in other ccTLDs,
%\eg \tld{ua} or \tld{sk}, also use Cloudflare but did not activate TLS~1.3 for
%some reason. This is not too likely as TLS~1.3 is enabled by default; however,

\begin{table*}
    \caption{Analysis of TLS-enabled domains with front-end by a major provider. Note that percentages are percentages of domains with successful TLS 1.x and
    TLS~1.3 handshakes, respectively, not just \textit{resolvable domains}. The special case of Akamai is discussed in Section~\ref{sub:active-scanning}.}
    \label{tab:tls-versions-over-groups}
    \small
    \centering
    \begin{tabular}{lrrrrrrrr}
        \toprule
        ~               &  \multicolumn{2}{c}{Alexa}                & \multicolumn{2}{c}{\cno}              & \multicolumn{2}{c}{gTLD}                    & \multicolumn{2}{c}{ccTLD} \\
        ~               & \%  TLS~1.3           & \% TLS 1.x            & \% TLS 1.3             & \% TLS 1.x         & \% TLS 1.3                   & \% TLS 1.x         & \% TLS 1.3            & \% TLS 1.x \\   
        \midrule
        Cloudflare      &  \tgb{59.8 (1)}  & \tgb{13.5 (1)}     & \tgb{35.4 (1)}   & \tgb{4.8 (2)}  & \tgb{70.6 (1)}         & \tgb{17.3 (1)} & \tgb{32.7 (1)}  & \tgb{3.4 (2)} \\
        Google          &  \tgb{11.3 (2)}  & \tgb{5.7  (3)}     & \tgb{1.4 (3)}    & 2.9 (6)       & 0.6 (5)                & 2.6 (5)       & 0.6 (5)         & 2.1 (5) \\
        Squarespace     &  \tgb{4.8 (3)}   & 1.0 (8)           & \tgb{29.8  (2)}  & 3.6 (4)       & \tgb{7.1 (2)}          & 1.7 (6)       & \tgb{5.5 (2)}   & 0.6 (6)\\
        Amazon          &   0.8 (4)        & \tgb{7.5 (2)}      & 0.6 (5)          & \tgb{4.3 (3)}  & \tgb{0.7 (3)}          & \tgb{7.6 (2)}  & 0.5 (6)         & \tgb{3.3 (3)} \\
        OVH             &   0.7 (5)        & 3.8 (4)          & 1.0 (4)          & 3.5 (5)       & 0.7 (4)                & 3.9 (4)      & \tgb{1.0 (3)}   & \tgb{5.8 (1)}  \\
        DigitalOcean    &   0.6 (6)        & 1.7 (6)           & 0.5 (6)          & 0.9 (7)       & 0.3 (6)                &  1.3 (7)      & 0.6 (4)         & 0.6 (7)\\
        Azure           &   0.1 (7)        & 1.3 (7)           & 0.0 (7)          & 0.4 (8)       & 0.0 (8)                & 0.3 (8)       & 0.0 (7)         & 0.3 (8)\\
        Alibaba         &   0.0 (8)        & 0.1 (9)           & 0.0 (8)          & 0.1 (9)       & 0.0 (7)                &  0.3 (9)      & 0.0 (8)         & 0.0 (9)\\
        GoDaddy         &   0.0 (9)        & 2.8 (5)           & 0 (9)            & \tgb{15.2 (1)} & 0 (9)                       & \tgb{6.4 (3)}  & 0.0                & 2.7 (4) \\
        \midrule
        (Akamai)        &   (0.0)             & (0.3\%)             & (0.0)               & (0.1)        & (0.0)                     &  (0.0)          & (0.0)              & (0.1)\\
        \bottomrule
    \end{tabular}
\end{table*}

% XHOSTING marker
\subsubsection{Impact of hosting services}
\label{sec:hostingimpact}

Table~\ref{tab:tls-versions-over-groups} provides an overview of TLS~1.3
deployment across our chosen domain groups. On the Alexa list, the biggest
player across all zones and groups is Cloudflare: 13.5\% of TLS-enabled domains
reside in their IP range, with Amazon (7.5\%) and Google (5.7\%) following at
some distance.  However, nearly 60\% of all TLS~1.3-enabled domains are hosted by
Cloudflare, and both Google and Amazon have much lower shares.

Cloudflare's dominance also extends to gTLDs, where more than 70\% of TLS
1.3-enabled domains are in Cloudflare's IP space. It is `just' over 30\% in \cno
and across the ccTLDs. Squarespace, interestingly, is not strongly
represented in most domain groups, except \cno---showing that the company hosts
many smaller sites not on the Alexa list. Similarly, GoDaddy generally has 
no TLS~1.3 deployment, but a number of domains in \cno and the
gTLDs host with them have. Amazon does not have significant deployment of TLS~1.3,
enabled. The VPS provider, OVH, rarely shows up with significant numbers,
except for ccTLDs. OVHs is reputedly a common choice among private customers
and smaller businesses who tend to host under their country's ccTLD.

We analyze the use of TLS~1.3 on Alexa domains in more detail. Previous
analyzing the deployment of a new security technology commonly showed
more high-ranking domains deploying the technology, especially if the new
mechanism had low risk to availability and little complexity. This is true, e.g.,
for Certificate Transparency and HSTS~\cite{Amann2017}, but also for
Certificate Authority Authorization~\cite{scheitle2018rise}. Our expectation is
hence that TLS~1.3 is also deployed more commonly on high-ranking sites.
However, our data does not support this: deployment is fairly
consistent across the entire range. TLS~1.3 has a
fraction of 21.7\% for the top 1M domains. For the top 1K, it is 21.9\%; for
the top 50K and 100K we find 27.3\% and 26.1\%, \ie a slight bump. For the top
500k, it falls to 22.2\% again. 

We filter our results to investigate the contributions of cloud providers
in the case of TLS~1.3-enabled domains in Table~\ref{tab:alexa-tls13-and-hosting}. 
Cloudflare is the dominant provider: of top 1M TLS~1.3-enabled domains, 59.8\%
are with Cloudflare. The percentage is even higher for domains in the middle
ranges of the Alexa list, reaching up to 82.9\%. Google is only relevant in the
Top 1K, where its share of TLS~1.3-enabled domains is 28.1\%---although its
market share rises slightly at the lower end of the Alexa ranks. Squarespace
host 4.8\% of all TLS~1.3 domains, but also with a market share shifted towards
the lower-ranking domains. Amazon has a relatively meager share across all ranks.

\begin{table*}
    \caption{TLS~1.3 deployment on Alexa domains. Percentages are given with respect
    to domains with successful TLS~1.3 handshakes. We omit Azure, Alibaba, and
    GoDaddy due to very low numbers.\label{tab:alexa-tls13-and-hosting}}
    \small
    \centering
    \begin{tabular}{lrrrrrr}
        \toprule
        TLS~1.3                   &  \% Top 1K    & \% Top 10K    & \%  Top 50K       &   Top 100K    & \% Top 500K       & \% Top 1M \\
        \midrule
        + Cloudflare        &  57.1         & 79.6          &    82.9           & 82.2          & 70.6              & 59.8 \\ 
        + Google            &  28.1         &  5.3          &     2.5           &  2.5          &  6.3              & 11.3 \\
        + Squarespace       &  0.5          &  0.2          &     0.1           &  0.3          &  1.8              & 4.8 \\
        + Amazon            &  1.0          &  0.4          &     0.5           &  0.6          &  0.8              & 0.8\\
        + OVH               &  0.0          &  0.4          &     0.3           &  0.4          &  0.7              & 0.7 \\
        + DigitalOcean      &  0.0          &  0.2          &     0.3           &  0.3          &  0.6              & 0.6 \\
%        + Azure             &  0.0          &  0.0          &     0.1           &  0.1          &  0.1              & 0.1 \\
        \midrule
        other setups        &  2.6          & 3.4           &   3.6             & 3.5           & 4.2               & 4.7 \\
        \bottomrule
    \end{tabular}
\end{table*}

\subsection{Use in research/education networks}
\label{sec:useinre}

In this section discuss results from the passive data
collections, using data from both the ICSI SSL Notary as well as our collection
effort in Australia.

\begin{figure}[t]
  \includegraphics[width=\columnwidth]{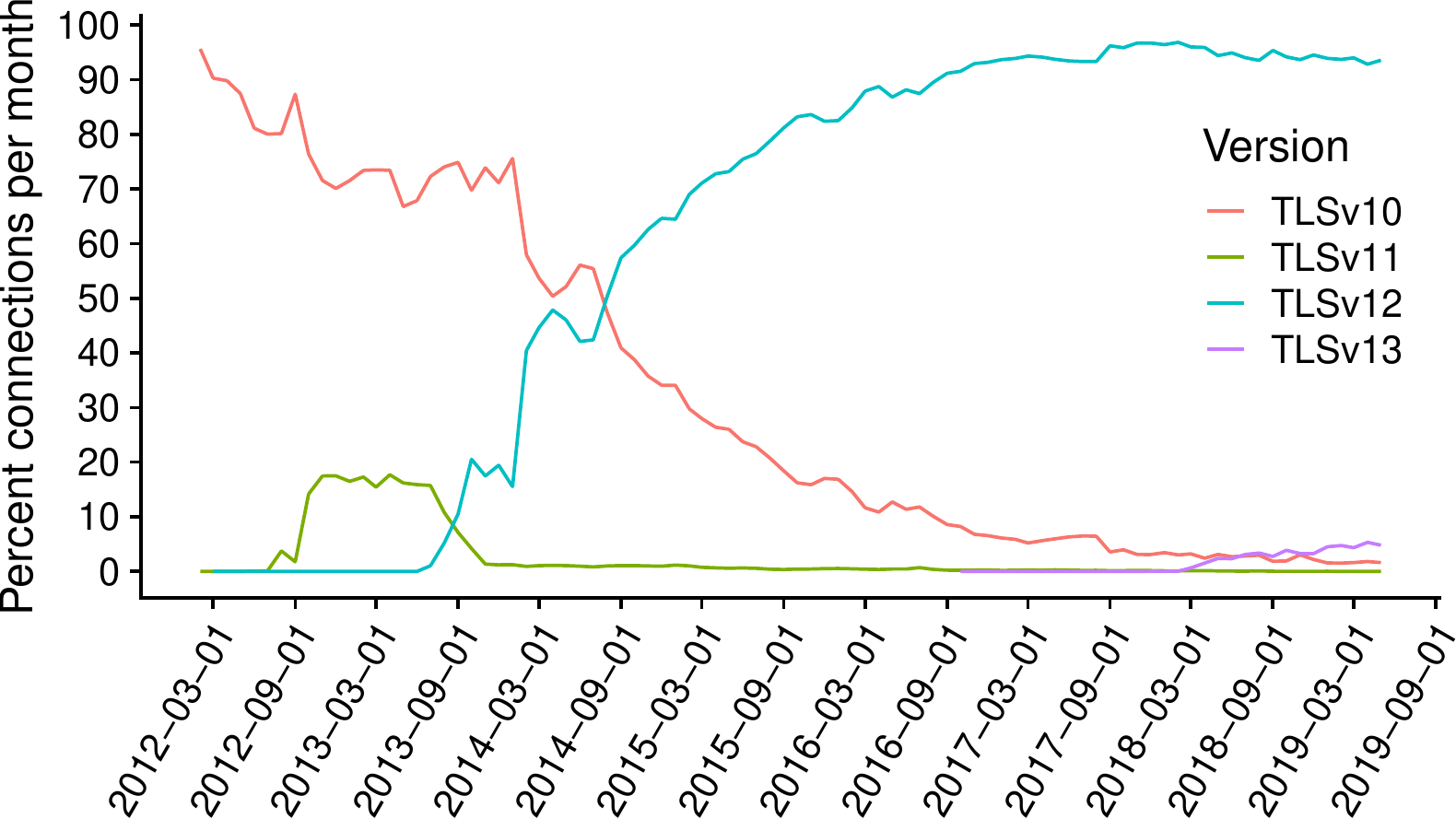}
	\caption{Negotiated TLS Versions since February 2012.}
  \label{fig:versions_negotiated}
	%\vspace{-0.5cm}
\end{figure}

\fref{fig:versions_negotiated} shows the versions of TLS that the ICSI Notary
saw being negotiated in our large-scale dataset since February 2012. To not
clutter the plot, we exclude SSLv2 and SSLv3, which did not see significant
use.

\begin{figure}[t]
  \includegraphics[width=\columnwidth]{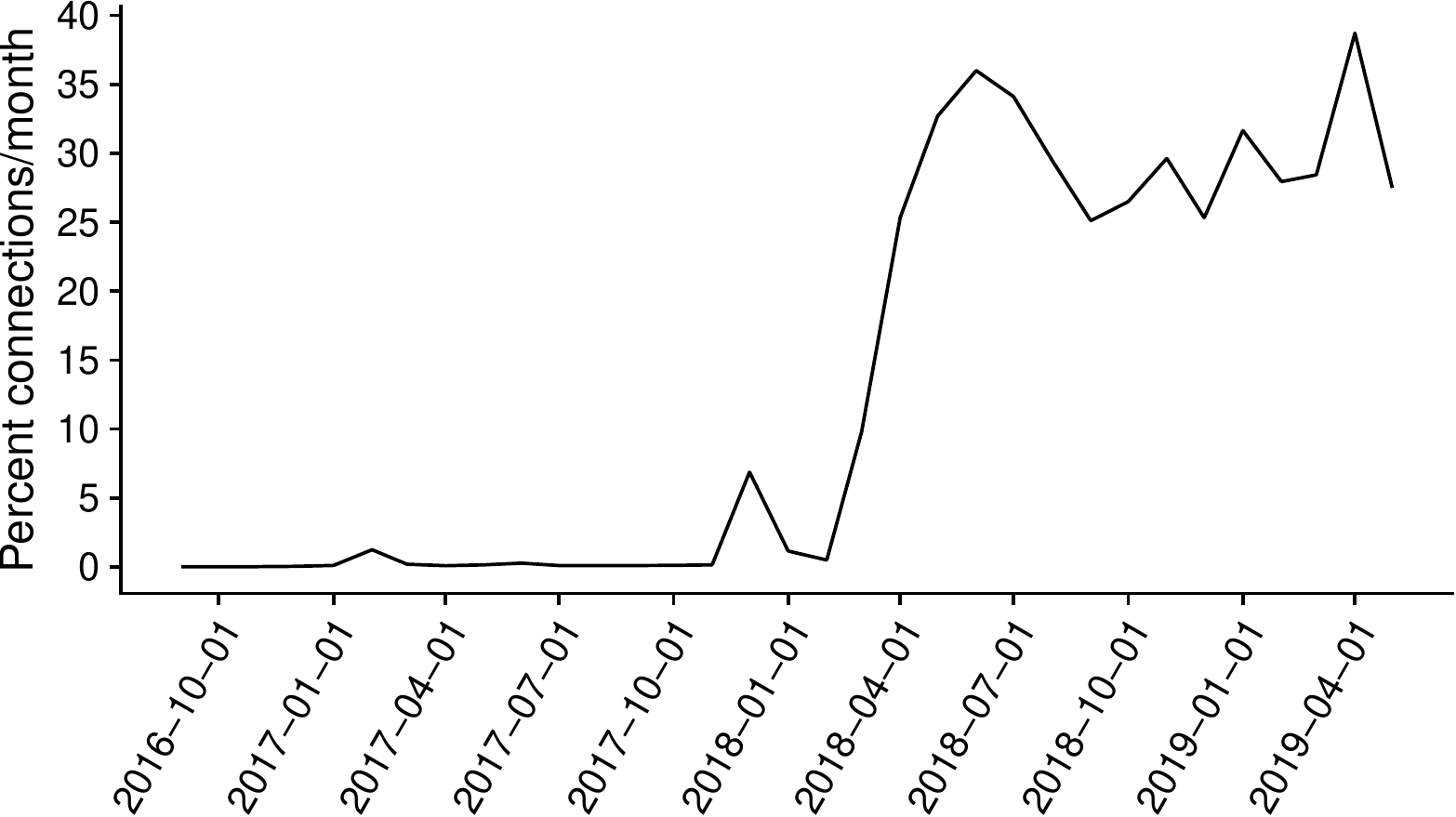}
	\caption{Client connections offering TLS~1.3.}
  \label{fig:versions_offered}
	%\vspace{-0.5cm}
\end{figure}

TLS was standardized in 2008. Yet, at the beginning of the graph in 2012, the
Notary saw basically zero use of TLS~1.2. Much software did not support TLS 1.2
at this point of time. For example, OpenSSL added support for TLS~1.2 in
version 1.0.1 (March 2012). The Notary did not see more than 50\% of
connections use TLS~1.2 before mid-2014. In contrast, TLS~1.3 is already seeing
a significant amount of use. As of April of this year
\negotiatedversionscategory{TLSv13}\% of connections negotiate some variant of
TLS~1.3. This is the case even though the RFC was only published in August
2018. \fref{fig:versions_offered} takes a look at client connections offering
TLS~1.3 in the notary data set. This gives an even more extreme picture---as of
April 2019, 39.8\% of clients advertise support for some variant of TLS~1.3.

% Note that for \fref{fig:versions_negotiated} the rise of TLS~1.2 support also
% coincides with both the Snowden revelations as well as attacks on earlier
% protocol versions. Kotzias \etal~\cite{kotzias2018coming} provide a thorough
% evaluation of these correlations.

Comparing the Notary figures with our data collection in Australia reveals that
TLS~1.3 traffic is significantly more commonly seen there;
\SYDnegotiatedversionscategory{TLSv13}\% of connections negotiate a variant of
TLS~1.3. This might be caused by different usage-patterns, which we show below.

\subsubsection{TLS~1.3 variants}
\label{sec:tlsvariants}

\todo{Maybe move to background?} With TLS~1.3, the way that the TLS protocol
version is negotiated changes significantly. In TLS~1.2 and below, the client
advertises the highest version it supports in the version field of the client
hello. The server selects the final version and returned it in the version
field of the server hello.

TLS~1.3 originally wanted to keep this approach. However, trials showed that
some servers did not react gracefully when exposed to version numbers greater
than 1.2~\cite{tls13-draft16}. Thus, with draft 16 of the TLS~1.3 RFC a new
approach was introduced. The client-hello always sends a version field
indicating TLS~1.2. A new \emph{supported versions} extension advertises a list
of versions that the client supports, hiding higher versions from non-TLS~1.3
servers.

Later, in draft 22, a similar approach was introduced for the server-hello,
after it was determined that middle-boxes also have problems with the new TLS
1.3 server hello (see \cite{tls13-draft22}).  Originally, TLS~1.3 wanted to
introduce a new, shorter server hello.  Instead, the final TLS~1.3 server hello
uses the exact same structure as the TLS~1.2 server hello and puts TLS~1.2
into its version field. If TLS~1.3 is negotiated, this version is put into the
supported versions extension, like on the client side. 

This change of advertising specific versions, instead of a maximum version,
also allows the negotiation of alternative versions of TLS~1.3. The
aforementioned number of \negotiatedversionscategory{TLSv13}\% negotiated
TLS~1.3 connections is actually split accross a number of different versions
that are negotiated using the supported versions extension.

\tref{tab:clientserverversionstable} shows the server negotiated versions as
well as the client offered versions that we observed during the month of April
2019. Note that for the client offered versions we use all the values present
in the supported versions extension; since several values can be present, the
total can exceed 100\%. If the supported versions extension is not sent
(pre-TLSv1.3 clients), the client version hello is used.

\begin{table}
\small\centering
\caption{Client-offered, and final negotiated TLS versions in
April 2019 at site $N1$.}
\begin{tabular}{@{}lrr@{}}
\emph{Version} & \emph{Server Conn.} & \emph{Client Conn.}\\
\toprule
TLS~10 & 1.83\% & \tg{33.33\%}\\
TLS~11 & 0.01\% & \tg{32.54\%}\\
TLS~12 & \tgb{93.6\%} & \tgb{84.69\%}\\
TLS~13 & 2.51\% & \tgb{34.4\%}\\
TLS~13-7E01 & none & $<0.01$\\
TLS~13-7E02 & none & $<0.01$\\
TLS~13-draft18 & none & 0.04\%\\
TLS~13-draft23 & 0.01\% & 0.36\%\\
TLS~13-draft26 & $<0.01$ & 0.01\%\\
TLS~13-draft27 & none & $<0.01$\\
TLS~13-draft28 & $<0.01$ & 0.02\%\\
TLS~13-FB23 & $<0.01$ & 0.01\%\\
TLS~13-FB26 & 2.05\% & 2.03\%\\
\bottomrule
\end{tabular}
\label{tab:clientserverversionstable}
\end{table}

The two TLS versions starting with FB are used exclusively by Facebook
services; the connections terminate at Facebook and Instagram servers. We
assume that these connections are mostly made by mobile apps. \todo{Why mention
then? Is stuff maybe different in the encrypted part?} We are not sure how
these connections differ from the final TLS~1.3 standard: for a passive
observer besides the fact that they use a different version number, they look
like typical TLS~1.3 connections.

% Verified in R notebook - to a passive observer the handshake messages look exactly
% the same.

We still see a few connections of two draft versions of TLS~1.3 being
negotiated (drafts 23, 26, and 28). These connections terminate  nearly
exclusively at Facebook servers; only  a few draft 23 and draft 28 connections
terminate at other servers (\eg \textit{gravatar}, \textit{atdmt.com}).

The 0x7E01 and 0x7E02 versions that were advertised by some clients are
experiments by Google Chrome to test slight changes to handshake behavior.  We
assume they are caused by old versions of Google Chrome still in circulation;
for the 0x7E01 case, we saw 3 connections (terminating at
\textit{gstatic.com}); for the 0x7E02 case we saw 16,548 connections to a mix
of domains, including Google, Facebook and a few smaller services.

Looking just at the clients that send the supported versions extension also
reveals that most clients also signal support for older versions of TLS in the
extension.  For clients sensing the extensions, 93.4\% of connections advertise
support for TLS~1.0 in it, 93.4\% for TLS~1.1 and 94.2\% for TLS~1.2. We are
not sure why this choice was made; servers supporting the supported versions
extension will have to support at least TLS~1.2 and some variant of TLS~1.3.
While an argument can be made to include TLS~1.2 in the list, there does not
seem to be a good reason to include even earlier versions.

The standardized TLS~1.3 is offered in 98.8\% of connections; the rest offers
one of the other TLS~1.3 variants. 56.7\% of
connections also contain one of the GREASE markers in the supported versions
fields. GREASE is a proposal by Google that introduces random numbers to some
field of the TLS handshake~\cite{greasedraft}. The goal is to expose bugs in software
that does not deal well with unknown values (which should be ignored).

Looking at the evolution of TLS versions offered by clients reveals that, especially
at the beginning, there were rapid changes. While in November 2016, virtually all observed TLS~1.3
connections advertised draft 16, this changed to all draft-18 by January 2018. Draft
18 in turn nearly completely disappears in Februrary 2018 - till then it was
responsible for nearly 100\% of TLS~1.3 connections. Starting in 2018 the situation
gets more complex with several different drafts, as well as proprietary versions of
Google and Facebook beind present simultaneously. Support for the final version of TLS~1.3 has
been growing quickly since October 2011.

% old text
% Taking a step back, we also examine temporal use of the different TLS~1.3 variants.
% \fref{fig:tls_1_3_negotiated_versions} shows all TLS~1.3 variants that we see used
% over time. As the plot shows, especially early on there are massive quick changes of
% deployments - versions see next to no use anymore within a few feeks; and while less
% pronounced it seems that these changes are still ongoing, e.g. with the Facebook
% versions of TLS~1.3. This is another datapoint that shows that at least the initial
% deployment of TLS~1.3 is driven and controlled by a small amount of entities.
% 
% \begin{figure}[t]
%   \includegraphics[width=\columnwidth]{tls_1_3_negotiated_versions}
% 	\caption{TLS~1.3 versions negotiated over time. \textbf{Note: If we use this plot, Johanna
% will make the labels nice. It is a bunch of work - so... only if we really need it}}
%   \label{fig:tls_1_3_negotiated_versions}
% \end{figure}

\subsubsection{Users of TLS~1.3}

Similar to Sec.~\ref{sec:hostingimpact}, we use IP ranges to determine hosting
providers and large services that are commonly seen hosting TLS services.
\tref{tab:tlsonethreeusetablebig} compares the use of TLS~1.3 and earlier TLS
versions of different services with each other.

\begin{table*}
\small\centering
	\caption{Percentage of connections and IP addresses speaking different TLS versions mapped
	to different providers.}
\begin{tabular}{@{}lrrrr|rrrr@{}}
	~ & \multicolumn{4}{c}{Notary Site $N1$} & \multicolumn{4}{c}{Australian University}\\
	~ & \multicolumn{2}{c}{\% Connections} & \multicolumn{2}{c}{\% IPs} & \multicolumn{2}{c}{\% Connections} & \multicolumn{2}{c}{\% IPs}\\
	~ & \emph{TLS1.3} & \emph{$\leq$ TLS1.2} & \emph{TLS1.3} & \emph{$\leq$ TLS1.2} & \emph{TLS1.3} & \emph{$\leq$ TLS1.2} & \emph{TLS1.3} & \emph{$\leq$ TLS1.2}\\
\toprule
	Facebook & \tgbs{60.78}\,(1) & 1.02\,(7) & 3.22\,(4) & 0.12\,(10) &			                 \tgbs{24.72}\,(2) & 1.54\,(6) & 1.15\,(5) & 0.12\,(11)\\
	Cloudflare & \tgs{9.66}\,(3) & 1.39\,(6) & \tgbs{70.45}\,(1) & 4.86\,(4) &			         5.79\,(4) & 1.04\,(7) & \tgbs{64.17}\,(1) & 4.44\,(3)\\
	Google & 8.33\,(4) & \tgs{12.87}\,(3) & 4.91\,(3) & 1.47\,(5) &			                     \tgbs{52.46}\,(1) & 6.75\,(4) & 5.33\,(3) & 2.52\,(5)\\
	Amazon & 0.42\,(5) & \tgbs{33.64}\,(2) & 2.32\,(5) & \tgbs{68.02}\,(1) &			           0.25\,(5) & \tgbs{31.23}\,(2) & 2.2\,(4) & \tgbs{44.18}\,(1)\\
	Akamai & 0.41\,(6) & 7.2\,(5) & 0.86\,(6) & 6.13\,(3) &			                             0.15\,(6) & 5.29\,(5) & 0.34\,(8) & 4\,(4)\\
	Digitalocean & 0.05\,(7) & 0.16\,(9) & 0.65\,(7) & 0.69\,(6) &			                     0.01\,(9) & 0.11\,(10) & 0.5\,(6) & 1.16\,(7)\\
	Squarespace & 0.04\,(8) & $<\!0.01$\,(12) & 0.01\,(11) & $<\!0.01$\,(12) &
	0.03\,(7) & $<0.01$\,(12) & 0.03\,(11) & $<0.01$\,(12)\\
	Alibaba & 0.02\,(9) & 0.04\,(11) & 0.04\,(10) & 0.04\,(11) &			                       $<0.01$\,(12) & 0.21\,(8) & 0.08\,(9) & 0.19\,(10)\\
	Ovh & 0.02\,(10) & 0.2\,(8) & 0.24\,(8) & 0.43\,(8) &			                               0.02\,(8) & 0.18\,(9) & 0.4\,(7) & 0.94\,(8)\\
	Azure & $<\!0.01$\,(11) & \tgs{8.88}\,(4) & 0.07\,(9) & 0.45\,(7) &			                 0.01\,(10) & 15.07\,(3) & 0.05\,(10) & 1.21\,(6)\\
	Godaddy & $<\!0.01$\,(12) & 0.06\,(10) & 0.01\,(12) & 0.37\,(9) &			                   $<0.01$\,(11) & 0.01\,(11) & 0.02\,(12) & 0.45\,(9)\\
\bottomrule
	Others & 20.26 (2) & 34.54 (1) & 17.2 (2) & 17.42 (2) & 16.56\,(3) & 38.58\,(1) & 25.74\,(2) & 40.78\,(2)\\
\end{tabular}
\label{tab:tlsonethreeusetablebig}
\end{table*}

Our data shows a few striking differences. First, at the moment an overwhelming
fraction of TLS~1.3 connection terminates at Facebook servers, while these only
make up a relatively small number of the IP addresses serving TLS~1.3. In
contrast, Cloudflare owns over 70\% of the IP addresses that we see serving
TLS~1.3. The difference to TLS~1.2 and earlier deployments is also large: while
more than 33\% of the TLS~1.2 and earlier connections terminate in Amazon IP
space, only 0.42\% of TLS~1.3 connections do so.

Comparing data with Australia reveals a few interesting differences; in Australia
traffic to Google and not to Facebook dominates. Cloudflare is similar, not being
responsible for a lot of connections, but owning a large amount of the TLS~1.3 IPs.

Both sites show that currently a small number of entities
are driving TLS~1.3 deployment. \tref{tab:tlsonethreeusetablebig} also shows
the striking difference of analyzing TLS connections by number of IP addresses
versus by number of connections, which gives a completely different view of the
ecosystem.

\subsubsection{Cipher use in TLS~1.3}

TLS~1.3 introduced cipher suites that are different from earlier versions,
currently limited to just 5 different cipher suites. Looking at the data in
April 2019, we see 128-bit AES in GCM mode dominating (79.2\% of connections);
we also see some use of 256-bit AES in GCM mode (14.4\%) as well as of
ChaCha20+Poly1305 (6.4\%). We see no use of the other 2 cipher suites (AES in
CCM mode). Looking at the different variants of TLS~1.3 does not change thie
result---128-bit AES in GCM mode is by far the most popular everywhere.

Comparing with earlier versions of TLS reveals a similar picture. 128-bit AES
in GCM mode with SHA256 and ECDHE with an RSA or ECDSA key exchange is used in
64.6\% of connections (51.9\% using RSA and 12.7\% using ECDSA key exchange).
Following this, we find 128-bit AES in GCM mode with SHA256 and ECDHE with an
RSA or ECDSA key exchange in 21.5\% of connections (20.1\% using RSA and 1.4\%
using ECDSA key exchange). The next most popular algorithm is 256-bit AES with
SHA384 in CBC mode with an RSA key exchange in 2.8\% of connections.

We did not find any cases where servers negotiated TLS~1.2 cipher suites (which
would break the specification in several ways).
%In TLS~1.2, cipher suites
%combine a key exchange, hash algorithm, symmetric cipher and key exchange
%whereas in TLS~1.3 they only specify the symmetric cipher and mode.
%
% \todo{Curves?}
% No curve data available :/ If we get accepted we can do this, but for this one
% it won't do. Another Bro analyzer problem that I did not notice till now.

\subsubsection{Resumed connections \& Early data}
\label{sec:earlydata}

In April connections of the ICSI Notary, 9.7\% of TLS~1.3 connections include
the pre-shared key extension (Australia: 12.6\%), indicating that client can resume a connection
with a server.\footnote{Another potential use-case is that the client and server
pre-negotiated an pre-shared-key out of band. This use-case is unusual.} In
92\% of cases (or 8.9\% of total connections; Australia: 96\% or 12.1\% of total), the server also replies with a
pre shared key extension which means that resumption (or PSK use) succeeds.

Besides session-resumption, TLS~1.3 also introduces a 0-RTT mode in which a
client can already send data in the first TCP packet. The client must have
connected to the server at an earlier time and must try to resume the
connection with already pre-established key material. The client can signal
that it wants to send 0-RTT data using the \emph{early data} extension. This
comes with certain drawbacks; using 0-RTT mode is vulnerable against replay
attacks, and applications must check for this.

6.8\% of TLS~1.3 connections, or 70\% of connections  in
which a client tries session resumption send the early data extension (Australia: 4.3\% of total or 33\%),
signaling that the client sent 0-RTT data.  Using our passive data we cannot
tell in how many cases the use of early data succeeds---the encryption is
already active at the point when a server signals if it did or did not accept
the early data.

Cloudflare published a blog post \cite{tls13-0rttCloudflare} in which they introduce 0-RTT
support for
their infrastructure. They estimate that around 40\% of connections might use
0-RTT in TLS~1.3 because about 40\% of TLS~1.2 connections use resumption.
These numbers are much larger than what we currently observe in practice.

The faster handshake of TLS~1.3 relies on the client already sending cryptographic
information in the first handshake packet. In some cases, the server will not accept
the chosen cryptographic parameters, prompting clients to re-send a second
client-hello using a hello retry request.
This happens in 4\% of connections (Australia; data not present in Notary collection).

\todo{Candidate for shortening}

\subsubsection{Other protocol features}

The TLS~1.3 connections we observe also use new extensions that have, to a
large degree, not been observed in earlier work.  69.1\% of April TLS~1.3
connections show support for the certificate compression extension, of which
only an outdated draft (authored by Google and Cloudflare employees)~\cite{certcompress}
exists.  13.0\% of April TLS~1.3 connections advertise support for the record
size limit extension~\cite{RFC8449},
which is intended for resource-limited clients. In addition, 78.3\% of April
TLS~1.3 connections support the signed certificate timestamp (SCT) extension.
SCTs are used together with the Certificate Transparency project~\cite{RFC6962,scheitle2018rise}. We also encountered 6
connections in all of 2019 that advertised support for the encrypted server
name indication.

Another commonly used extension is \emph{application layer next protocol
negotiation}. This extension is, for example, used to negotiate HTTP/2. It is
sent by the client in 96.1\% of TLS~1.3 connections in April 2019. In 20.5\% of
cases clients signal that they just support HTTP~1.1. 55.1\% of connections
signal support for HTTP~2 and 1.1. Surprisingly, four of these connections list
HTTP~1.1 first, signaling that they prefer it over HTTP2. 3.8\% of connections
just advertise support for HTTP~2; these clients probably will still support
HTTP~1.1 if the server does not support the extension.The remaining 24.0\% of
connections signal support for SPDY~3.1, 3.0 and HTTP~1.1 (but not HTTP~2). On
the server side, the extension is encrypted, so we cannot tell what TLS~1.3
servers select.

\subsection{Use in the Android ecosystem}

As of today, Android does not provide native TLS~1.3 support
to Android apps. Only beta versions of
Android Q do so since March 2019~\cite{androidQfeatures}. However, as
Razaghpanah \etal{} demonstrated~\cite{razaghpanah2017studying}, most
Android apps use native OS libraries with default configurations 
for their TLS needs. This implies that only a very small fraction of Android users
running beta versions of Android Q, as well as those users who have installed 
apps developed by large companies like Google, Facebook, and Mozilla can use 
TLS~1.3 in their handsets. In this section, we use Lumen data to look at the 
deployment of TLS~1.3 in the Android ecosystem, both at the client- and at 
the server-side.

\subsubsection{Client-side vs. server-side support}

The fraction of Lumen-captured TLS~1.3 connections has gone up from 0.01\% 
of all TLS connections in January of 2018 to 4\% in March 2019. However,
TLS~1.3 support in Android apps is lagging behind TLS~1.2 support 
which still accounts for 94.9\% of observed TLS connections in March 2019.

\begin{figure}[t]
  \includegraphics[width=\columnwidth]{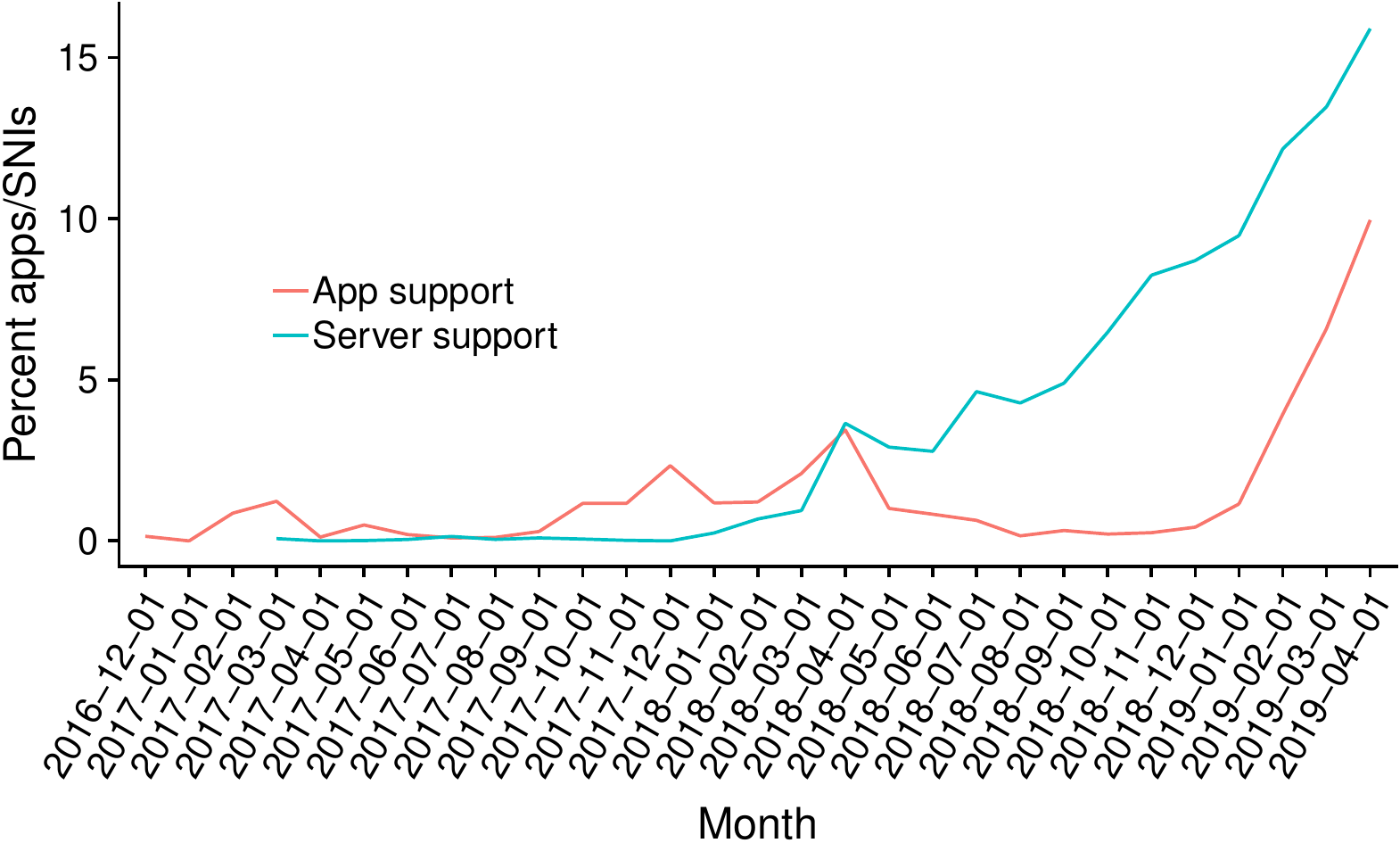}
	\caption{Percentage of mobile apps and servers offering TLS~1.3.}
  \label{fig:lumen_app_server_support_for_tlsv13}
\end{figure}

Examining TLS~1.3 deployment and support on a per-application basis, shows 
interesting dynamics and clearly identifies the main actors driving this effort. 
Figure~\ref{fig:lumen_app_server_support_for_tlsv13} shows how Android app
support for TLS~1.3 has evolved over time compared to server-side support.  
A server is marked as supporting TLS~1.3 when it either negotiated a TLS~1.3
connection, or when it set the appropriate downgrade marker indicating TLS~1.3
support.
%
%\todo{NARSEO: The TLS~1.2 downgrade reads weird to me. Can we clarify that those were
%1.3 CH but they negotiated 1.2 in the end? Also, how can we determine if the
%server does support if if it did downgrade it?}
%
The figure demonstrates how client-support lags behind server-side for
Android apps. The sharp increase for app support of TLS~1.3 is
partially caused by early users of the Android Q beta (March 2019).

Prior to the release of this beta, apps that advertise support for TLS~1.3 do 
so using their own TLS framework or external open-source TLS libraries like Facebook's 
Fizz~\cite{fbFizz}, or OpenSSL 1.1. Specifically, Lumen data shows that beyond major 
Android browsers known to support TLS~1.3 ---\eg beta versions of Chromium and Firefox
and other browsers based on them, which are not present in Lumen's dataset to preserve users'
anonymity---, only a handful of apps supported TLS~1.3 before it became standardized. 
The Facebook family of apps have implemented draft versions of TLS~1.3 since early 2017, 
replacing the implementaions of older drafts with new ones as they were being proposed for
standardization. This indicates that, due to the absence of native OS support, early 
TLS~1.3 support in Android has been limited to apps developed by companies with large 
development teams. 

The TLS~1.3 draft versions that have experienced the most dramatic increases and subsequent 
decreases in use are the ones deployed by companies like Facebook and Google. These companies 
have a privileged position thanks to their control over both the client- and the server-side. 
For instance, Facebook's TLS~1.3 custom draft version 23 sees a sharp decline in the Spring of 2018, 
going down from 49\% of all negotiated TLS~1.3 versions in April of 2018, to 0.1\% the next 
month, replaced by Facebook's custom draft version 26, which accounted for 47\% of all negotiated 
TLS~1.3 versions in May of 2018. 
We stress, however, that Facebook's family of apps do not advertise Facebook's custom versions 
of TLS~1.3 as the one with the highest priority in the supported versions extension. We speculate 
that this is done to avoid breaking TLS~1.3 servers that don't support Facebook's own versions of 
TLS~1.3.
%which results in Facebook's 
%versions not appearing in this figure. 
%This is in line with our observations in~\xref{sec:tlsvariants}. 
%
The final TLS~1.3 standard has been rapidly adopted by those Android apps that 
had previosly supported drafts of the protocol. This is evident by examining both
advertised and negotiated versions. However, given application developers' reliance 
of platform-provided TLS libraries, it seems unlikely that we will observe a massive 
support of TLS~1.3 in Android applications until Android Q is officially released 
in the second-half of 2019.

%\todo{RH: Why not?}
%Abbas: I don't know. I think to avoid breaking servers that don't support it?

%\todo{This is not saying much. Can we be more specific? Is this rapid adoption
%driven by apps running on Beta versions of Android Q?}

%\todo{We have a good bunch of figures that could be verbosely discussed in the
%text. That could be useful to add more content / material for a long paper :)}

%\todo{Even before Android Q got publicly released, there seems to be an increase 
%in the \% of apps supporting TLS~1.3 around January. Who are they?}

%\begin{figure}[t]
%  \includegraphics[width=\columnwidth]{servers_supporting_tlsv13_temporal}
%	\caption{Server-side support for TLS-1.3 in the Lumen dataset.}
%  \label{fig:lumen_server_support_for_tlsv13}
%\end{figure}

%\begin{figure}[t]
%  \includegraphics[width=\columnwidth]{clients_supporting_tlsv13_temporal}
%	\caption{Percentage of mobile app packages offering TLS-1.3.}
%  \label{fig:lumen_app_support_for_tlsv13}
%\end{figure}

%\begin{figure}[t]
%  \includegraphics[width=\columnwidth]{negotiated_tlsv13_version_breakdown_temporal}
%	\caption{Breakdown of different TLS-1.3 versions negotiated in Android over time.}
%  \label{fig:lumen_negotiated_tlsv13_versions}
%\end{figure}

%\begin{figure}[t]
%  \includegraphics[width=\columnwidth]{advertised_tlsv13_version_breakdown_temporal}
%	\caption{Breakdown of the highest version of TLS-1.3 versions advertised in Android over time.}
%  \label{fig:lumen_advertised_tlsv13_versions}
%\end{figure}

\section{Discussion}

The adoption of TLS 1.3 seems to be
happening much faster than ever before for a new TLS version. Growing awareness
of security and privacy may play some role, but our findings identify a
different primary reason: a small set of cloud providers host a large number of
domains and is able to activate the new protocol on their behalf. This
corresponds with early support in mainstream browsers, at least outside of the
mobile world. Our data is quite consistent in this regard: although deployment
among DNS zones vary quite dramatically, when TLS 1.3 is supported, it is most
commonly because of Google, Facebook, and Cloudflare.

Facebook is a particularly interesting case as they deployed Facebook-specific
versions and experimented with them. Control of both endpoints---an app on the
mobile device, and the Facebook server farms---are the necessary ingredients
here. A similar statement can be made about Google, who control the Chrome
browser and a number of the most popular Internet services, and who have
contributed to the protocol development and detected the problematic cases of
malfunctioning middleboxes.

Cloudflare is the front-end provider responsible for most TLS 1.3 handshakes in
our active scans---for Alexa domains, Cloudflare accounts for 60\% of TLS
1.3-enabled domains. Smaller hosters complement this picture, like Squarespace
with their respectable share of Alexa domains and ccTLD domains. We have found
evidence that, in some countries, similar providers have the same role.

On the face of it, one could thus speak of a net benefit for consumers.
However, our data also shows that uptake of the new protocol version is
remarkably different outside the ecosystem of the companies that drove the
development of TLS 1.3---as in \cno and most ccTLDs, especially the important
zones \tld{de} and \tld{fr}, which have a very low adoption rate. While the big
providers profit from their ability to test-drive new protocols in many
variants, providers like Amazon or Azure control only one endpoint and have a
competitive disadvantage when developing a roll-out strategy. In the history of
TLS, this is an unprecedented situation. In our view, the question will be
whether the current market concentration of TLS 1.3 will attract more customers
to the big providers---to their Web services and clients alike---or whether the
playing field will be level eventually. We note that similar questions surround
other network protocols like QUIC.

TLS 1.3 has a strong, positive impact on privacy. While the server name is not
(yet) encrypted, nearly everything else of importance is. This makes it
difficult to deploy passive measurements to understand issues with new
protocols. Compared to previous studies on TLS 1.2, we were already able to
gather less data about TLS 1.3 details. Active scans can help to some degree,
but they have an impact on the Internet's server population and they are not a
good method to test the many varieties of TLS 1.3 and future protocols that may
be test-driven in the field before standardization.  In fact, the
aforementioned providers are once again in the best position to understand the
development of new security protocols. Although they seem keen at this stage to
share their insights with the research and development community, this may
collide with business interests at some points. It seems worthwhile to think
about methods of collaboration to develop new protocols.

One interesting lesson from the deployment of TLS~1.3 is the difficulty of
introducing this new version---even though TLS had version numbers, middleboxes and
server software did not cope well with higher version numbers, triggering several
protocol redesigns. Google tries to prevent this from happening again by introducing
random numbers to a lot of handshake elements (like the supported version extension)
- it will be interesting to see if this approach succeeds.

\section{Reproducible Research}

We wish to support other researchers repeating, replicating, or reproducing our
measurements. We publish all tools used in the preparation and execution of our
active scans and the code for each analysis step. We will release our data of
active scans in both raw (PCAP traces) and processed format (CSV). We cannot
release data from passive monitoring or Lumen collection for ethical and legal
reasons.

\section{Conclusion}

We presented a first study of deployment and use of TLS 1.3, including use in
mobile applications. Our key finding is that TLS 1.3 has considerable
deployment already. However, this is linked to strong market concentration:
very few providers like Cloudflare control a large number of domains and roll
out the new protocol.  Deployment elsewhere is strongly lagging behind.  In the
mobile ecosystem, Google and Facebook are the dominant users. In our passively
obtained data, we also observe that TLS 1.3 is mostly used in connections that
terminate at servers of these two companies, highlighting their importance for
consumers.

We will monitor how TLS~1.3 deployment will change in the future with the adoption
of OpenSSL 1.1 in more Linux distributions. This will also give us a chance to
measure the update patterns of a large part of the user-visible Internet.

\bibliographystyle{./acmart-master/ACM-Reference-Format}
{
% \small checker complains about this
\renewcommand*{\bibliofont}{\normalsize}
\bibliography{paper}

%%% -*-BibTeX-*-
%%% Do NOT edit. File created by BibTeX with style
%%% ACM-Reference-Format-Journals [18-Jan-2012].

\begin{thebibliography}{62}

%%% ====================================================================
%%% NOTE TO THE USER: you can override these defaults by providing
%%% customized versions of any of these macros before the \bibliography
%%% command.  Each of them MUST provide its own final punctuation,
%%% except for \shownote{}, \showDOI{}, and \showURL{}.  The latter two
%%% do not use final punctuation, in order to avoid confusing it with
%%% the Web address.
%%%
%%% To suppress output of a particular field, define its macro to expand
%%% to an empty string, or better, \unskip, like this:
%%%
%%% \newcommand{\showDOI}[1]{\unskip}   % LaTeX syntax
%%%
%%% \def \showDOI #1{\unskip}           % plain TeX syntax
%%%
%%% ====================================================================

\ifx \showCODEN    \undefined \def \showCODEN     #1{\unskip}     \fi
\ifx \showDOI      \undefined \def \showDOI       #1{#1}\fi
\ifx \showISBNx    \undefined \def \showISBNx     #1{\unskip}     \fi
\ifx \showISBNxiii \undefined \def \showISBNxiii  #1{\unskip}     \fi
\ifx \showISSN     \undefined \def \showISSN      #1{\unskip}     \fi
\ifx \showLCCN     \undefined \def \showLCCN      #1{\unskip}     \fi
\ifx \shownote     \undefined \def \shownote      #1{#1}          \fi
\ifx \showarticletitle \undefined \def \showarticletitle #1{#1}   \fi
\ifx \showURL      \undefined \def \showURL       {\relax}        \fi
% The following commands are used for tagged output and should be
% invisible to TeX
\providecommand\bibfield[2]{#2}
\providecommand\bibinfo[2]{#2}
\providecommand\natexlab[1]{#1}
\providecommand\showeprint[2][]{arXiv:#2}

\bibitem[\protect\citeauthoryear{??}{and}{[n.d.]}]%
        {androidQfeatures}
 \bibinfo{year}{[n.d.]}\natexlab{}.
\newblock \bibinfo{title}{{Android Q features and APIs}}.
\newblock
\newblock
\newblock
\shownote{\url{https://developer.android.com/preview/features\#tls-1.3}.}


\bibitem[\protect\citeauthoryear{??}{zee}{[n.d.]}]%
        {zeek}
 \bibinfo{year}{[n.d.]}\natexlab{}.
\newblock \bibinfo{title}{Zeek Network Security Monitor}.
\newblock
\newblock
\newblock
\shownote{\url{https://www.zeek.org/}.}


\bibitem[\protect\citeauthoryear{{A. Ghedini and V. Vasiliev}}{{A. Ghedini and
  V. Vasiliev}}{2019}]%
        {certcompress}
\bibfield{author}{\bibinfo{person}{{A. Ghedini and V. Vasiliev}}.}
  \bibinfo{year}{2019}\natexlab{}.
\newblock \bibinfo{title}{{TLS Certificate Compression}}.
\newblock
\newblock
\newblock
\shownote{\url{https://tools.ietf.org/html/draft-ietf-tls-certificate-compression-05}.}


\bibitem[\protect\citeauthoryear{Adrian, Bhargavan, Durumeric, Gaudry, Green,
  Halderman, Heninger, Springall, Thom{\'e}, Valenta, VanderSloot, Wustrow,
  Zanella-B{\'e}guelin, and Zimmermann}{Adrian et~al\mbox{.}}{2015}]%
        {logjam}
\bibfield{author}{\bibinfo{person}{David Adrian}, \bibinfo{person}{Karthikeyan
  Bhargavan}, \bibinfo{person}{Zakir Durumeric}, \bibinfo{person}{Pierrick
  Gaudry}, \bibinfo{person}{Matthew Green}, \bibinfo{person}{J.~Alex
  Halderman}, \bibinfo{person}{Nadia Heninger}, \bibinfo{person}{Drew
  Springall}, \bibinfo{person}{Emmanuel Thom{\'e}}, \bibinfo{person}{Luke
  Valenta}, \bibinfo{person}{Benjamin VanderSloot}, \bibinfo{person}{Eric
  Wustrow}, \bibinfo{person}{Santiago Zanella-B{\'e}guelin}, {and}
  \bibinfo{person}{Paul Zimmermann}.} \bibinfo{year}{2015}\natexlab{}.
\newblock \showarticletitle{Imperfect Forward Secrecy: How {D}iffie-{H}ellman
  Fails in Practice}. In \bibinfo{booktitle}{\emph{Proc. ACM SIGSAC Conference
  on Computer and Communications Security (CCS)}}.
\newblock


\bibitem[\protect\citeauthoryear{Akamai}{Akamai}{2019}]%
        {akamaiTLS}
\bibfield{author}{\bibinfo{person}{Akamai}.} \bibinfo{year}{2019}\natexlab{}.
\newblock \bibinfo{title}{{TLS 1.3 support is coming this spring}}.
\newblock
  \bibinfo{howpublished}{\url{https://blogs.akamai.com/2018/04/tls-13-this-spring.html
  }}.
\newblock


\bibitem[\protect\citeauthoryear{Akhawe, Amann, Vallentin, and Sommer}{Akhawe
  et~al\mbox{.}}{2013}]%
        {sslerrors}
\bibfield{author}{\bibinfo{person}{D. Akhawe}, \bibinfo{person}{J. Amann},
  \bibinfo{person}{M. Vallentin}, {and} \bibinfo{person}{R. Sommer}.}
  \bibinfo{year}{2013}\natexlab{}.
\newblock \showarticletitle{{Here's My Cert, So Trust Me, Maybe?: Understanding
  {TLS} Errors on the Web}}. In \bibinfo{booktitle}{\emph{Proc. of the
  International Web Conference (WWW)}}.
\newblock


\bibitem[\protect\citeauthoryear{AlFardan and Paterson}{AlFardan and
  Paterson}{2013}]%
        {lucky13}
\bibfield{author}{\bibinfo{person}{N.~J. AlFardan} {and} \bibinfo{person}{K.~G.
  Paterson}.} \bibinfo{year}{2013}\natexlab{}.
\newblock \showarticletitle{Lucky Thirteen: Breaking the {TLS} and {DTLS}
  Record Protocols}. In \bibinfo{booktitle}{\emph{Proc. IEEE Symposium on
  Security and Privacy (S\&P)}}.
\newblock


\bibitem[\protect\citeauthoryear{Amann, Gasser, Scheitle, Brent, Carle, and
  Holz}{Amann et~al\mbox{.}}{2017a}]%
        {amann2017mission}
\bibfield{author}{\bibinfo{person}{Johanna Amann}, \bibinfo{person}{Oliver
  Gasser}, \bibinfo{person}{Quirin Scheitle}, \bibinfo{person}{Lexi Brent},
  \bibinfo{person}{Georg Carle}, {and} \bibinfo{person}{Ralph Holz}.}
  \bibinfo{year}{2017}\natexlab{a}.
\newblock \showarticletitle{Mission accomplished?: {HTTPS} security after
  diginotar}. In \bibinfo{booktitle}{\emph{Proceedings of the 2017 Internet
  Measurement Conference}}. ACM, \bibinfo{pages}{325--340}.
\newblock


\bibitem[\protect\citeauthoryear{Amann, Gasser, Scheitle, Brent, Carle, and
  Holz}{Amann et~al\mbox{.}}{2017b}]%
        {Amann2017}
\bibfield{author}{\bibinfo{person}{J. Amann}, \bibinfo{person}{O. Gasser},
  \bibinfo{person}{Q. Scheitle}, \bibinfo{person}{L. Brent},
  \bibinfo{person}{G. Carle}, {and} \bibinfo{person}{R. Holz}.}
  \bibinfo{year}{2017}\natexlab{b}.
\newblock \showarticletitle{Mission accomplished? {HTTPS} security after
  {DigiNotar}}. In \bibinfo{booktitle}{\emph{Proc. ACM Int. Measurement
  Conference (IMC)}}. \bibinfo{address}{London}.
\newblock


\bibitem[\protect\citeauthoryear{Amann, Sommer, Vallentin, and Hall}{Amann
  et~al\mbox{.}}{2013}]%
        {noattacknecessary}
\bibfield{author}{\bibinfo{person}{Johanna Amann}, \bibinfo{person}{Robin
  Sommer}, \bibinfo{person}{Matthias Vallentin}, {and} \bibinfo{person}{Seth
  Hall}.} \bibinfo{year}{2013}\natexlab{}.
\newblock \showarticletitle{{No Attack Necessary: The Surprising Dy\-na\-mi\-cs
  of {SSL} Trust Relationships}}. In \bibinfo{booktitle}{\emph{Proc. Annual
  Computer Security Applications Conference}}.
\newblock


\bibitem[\protect\citeauthoryear{Amann, Vallentin, Hall, and Sommer}{Amann
  et~al\mbox{.}}{2012}]%
        {notarypaper}
\bibfield{author}{\bibinfo{person}{J. Amann}, \bibinfo{person}{M. Vallentin},
  \bibinfo{person}{S. Hall}, {and} \bibinfo{person}{R. Sommer}.}
  \bibinfo{year}{2012}\natexlab{}.
\newblock \bibinfo{booktitle}{\emph{{Extracting Certificates from Live Traffic:
  A Near Real-Time {SSL} Notary Service}}}.
\newblock \bibinfo{type}{{T}echnical {R}eport} TR-12-014.
  \bibinfo{institution}{ICSI}.
\newblock


\bibitem[\protect\citeauthoryear{Aviram, Schinzel, Somorovsky, Heninger,
  Dankel, Steube, Valenta, Adrian, Halderman, Dukhovni, K{\"a}sper, Cohney,
  Engels, Paar, and Shavitt}{Aviram et~al\mbox{.}}{2016}]%
        {197245}
\bibfield{author}{\bibinfo{person}{Nimrod Aviram}, \bibinfo{person}{Sebastian
  Schinzel}, \bibinfo{person}{Juraj Somorovsky}, \bibinfo{person}{Nadia
  Heninger}, \bibinfo{person}{Maik Dankel}, \bibinfo{person}{Jens Steube},
  \bibinfo{person}{Luke Valenta}, \bibinfo{person}{David Adrian},
  \bibinfo{person}{J.~Alex Halderman}, \bibinfo{person}{Viktor Dukhovni},
  \bibinfo{person}{Emilia K{\"a}sper}, \bibinfo{person}{Shaanan Cohney},
  \bibinfo{person}{Susanne Engels}, \bibinfo{person}{Christof Paar}, {and}
  \bibinfo{person}{Yuval Shavitt}.} \bibinfo{year}{2016}\natexlab{}.
\newblock \showarticletitle{{DROWN: Breaking {TLS} Using {SSLv2}}}. In
  \bibinfo{booktitle}{\emph{Proc. USENIX Security Symposium}}.
\newblock


\bibitem[\protect\citeauthoryear{Badertscher, Matt, Maurer, Rogaway, and
  Tackmann}{Badertscher et~al\mbox{.}}{2015}]%
        {badertscher2015augmented}
\bibfield{author}{\bibinfo{person}{Christian Badertscher},
  \bibinfo{person}{Christian Matt}, \bibinfo{person}{Ueli Maurer},
  \bibinfo{person}{Phillip Rogaway}, {and} \bibinfo{person}{Bj{\"o}rn
  Tackmann}.} \bibinfo{year}{2015}\natexlab{}.
\newblock \showarticletitle{Augmented secure channels and the goal of the {TLS}
  1.3 record layer}. In \bibinfo{booktitle}{\emph{International Conference on
  Provable Security}}. Springer, \bibinfo{pages}{85--104}.
\newblock


\bibitem[\protect\citeauthoryear{Bellare and Tackmann}{Bellare and
  Tackmann}{2016}]%
        {bellare2016multi}
\bibfield{author}{\bibinfo{person}{Mihir Bellare} {and}
  \bibinfo{person}{Bj{\"o}rn Tackmann}.} \bibinfo{year}{2016}\natexlab{}.
\newblock \showarticletitle{The multi-user security of authenticated
  encryption: {AES-GCM} in {TLS} 1.3}. In \bibinfo{booktitle}{\emph{Annual
  International Cryptology Conference}}. Springer.
\newblock


\bibitem[\protect\citeauthoryear{Benjamin}{Benjamin}{2019}]%
        {greasedraft}
\bibfield{author}{\bibinfo{person}{D. Benjamin}.}
  \bibinfo{year}{2019}\natexlab{}.
\newblock \bibinfo{title}{{Applying GREASE to TLS Extensibility}}.
\newblock
\newblock
\urldef\tempurl%
\url{https://tools.ietf.org/html/draft-ietf-tls-grease-02}
\showURL{%
\tempurl}


\bibitem[\protect\citeauthoryear{Beurdouche, Delignat-Lavaud, Kobeissi,
  Pironti, and Bhargavan}{Beurdouche et~al\mbox{.}}{2015}]%
        {beurdouche2015flextls}
\bibfield{author}{\bibinfo{person}{Benjamin Beurdouche},
  \bibinfo{person}{Antoine Delignat-Lavaud}, \bibinfo{person}{Nadim Kobeissi},
  \bibinfo{person}{Alfredo Pironti}, {and} \bibinfo{person}{Karthikeyan
  Bhargavan}.} \bibinfo{year}{2015}\natexlab{}.
\newblock \showarticletitle{$\{$FLEXTLS$\}$: A Tool for Testing $\{$TLS$\}$
  Implementations}. In \bibinfo{booktitle}{\emph{9th $\{$USENIX$\}$ Workshop on
  Offensive Technologies ($\{$WOOT$\}$ 15)}}.
\newblock


\bibitem[\protect\citeauthoryear{Bhargavan, Blanchet, and Kobeissi}{Bhargavan
  et~al\mbox{.}}{2017a}]%
        {bhargavan2017verified}
\bibfield{author}{\bibinfo{person}{Karthikeyan Bhargavan},
  \bibinfo{person}{Bruno Blanchet}, {and} \bibinfo{person}{Nadim Kobeissi}.}
  \bibinfo{year}{2017}\natexlab{a}.
\newblock \showarticletitle{Verified models and reference implementations for
  the {TLS} 1.3 standard candidate}. In \bibinfo{booktitle}{\emph{Proc. IEEE
  Symposium on Security and Privacy (S\&P)}}. IEEE.
\newblock


\bibitem[\protect\citeauthoryear{Bhargavan, Delignat-Lavaud, Fournet,
  Kohlweiss, Pan, Protzenko, Rastogi, Swamy, Zanella-B{\'e}guelin, and
  Zinzindohou{\'e}}{Bhargavan et~al\mbox{.}}{2017b}]%
        {bhargavan2017implementing}
\bibfield{author}{\bibinfo{person}{Karthikeyan Bhargavan},
  \bibinfo{person}{Antoine Delignat-Lavaud}, \bibinfo{person}{C{\'e}dric
  Fournet}, \bibinfo{person}{Markulf Kohlweiss}, \bibinfo{person}{Jianyang
  Pan}, \bibinfo{person}{Jonathan Protzenko}, \bibinfo{person}{Aseem Rastogi},
  \bibinfo{person}{Nikhil Swamy}, \bibinfo{person}{Santiago
  Zanella-B{\'e}guelin}, {and} \bibinfo{person}{Jean Zinzindohou{\'e}}.}
  \bibinfo{year}{2017}\natexlab{b}.
\newblock \showarticletitle{Implementing and proving the {TLS} 1.3 record
  layer}. In \bibinfo{booktitle}{\emph{Proc. IEEE Symposium on Security and
  Privacy (S\&P)}}.
\newblock


\bibitem[\protect\citeauthoryear{Chuat, Szalachowski, Perrig, Laurie, and
  Messeri}{Chuat et~al\mbox{.}}{2015}]%
        {ctgossip}
\bibfield{author}{\bibinfo{person}{L. Chuat}, \bibinfo{person}{P.
  Szalachowski}, \bibinfo{person}{A. Perrig}, \bibinfo{person}{B. Laurie},
  {and} \bibinfo{person}{E. Messeri}.} \bibinfo{year}{2015}\natexlab{}.
\newblock \showarticletitle{{Efficient Gossip Protocols for Verifying the
  Consistency of Certificate Logs}}. In \bibinfo{booktitle}{\emph{2015 IEEE
  Conference on Communications and Network Security (CNS)}}.
\newblock
\urldef\tempurl%
\url{https://doi.org/10.1109/CNS.2015.7346853}
\showDOI{\tempurl}


\bibitem[\protect\citeauthoryear{Clark and van Oorschot}{Clark and van
  Oorschot}{2013}]%
        {clark2013sok}
\bibfield{author}{\bibinfo{person}{J. Clark} {and} \bibinfo{person}{P. van
  Oorschot}.} \bibinfo{year}{2013}\natexlab{}.
\newblock \showarticletitle{{SoK}: {SSL} and {HTTPS}: Revisiting past
  challenges and evaluating certificate trust model enhancements}. In
  \bibinfo{booktitle}{\emph{Proc. IEEE Symposium on Security and Privacy
  (S\&P)}}.
\newblock


\bibitem[\protect\citeauthoryear{Coat}{Coat}{2017}]%
        {bluecoatTLS13}
\bibfield{author}{\bibinfo{person}{Blue Coat}.}
  \bibinfo{year}{2017}\natexlab{}.
\newblock \bibinfo{title}{{ProxySG, ASG and WSS will interrupt SSL connections
  when clients using TLS 1.3 access sites also using TLS 1.3}}.
\newblock
  \bibinfo{howpublished}{\url{http://bluecoat.force.com/knowledgebase/articles/Technical_Alert/000032878}}.
\newblock


\bibitem[\protect\citeauthoryear{Cremers, Horvat, Hoyland, Scott, and van~der
  Merwe}{Cremers et~al\mbox{.}}{2017}]%
        {cremers2017comprehensive}
\bibfield{author}{\bibinfo{person}{Cas Cremers}, \bibinfo{person}{Marko
  Horvat}, \bibinfo{person}{Jonathan Hoyland}, \bibinfo{person}{Sam Scott},
  {and} \bibinfo{person}{Thyla van~der Merwe}.}
  \bibinfo{year}{2017}\natexlab{}.
\newblock \showarticletitle{A comprehensive symbolic analysis of {TLS} 1.3}. In
  \bibinfo{booktitle}{\emph{Proc. ACM SIGSAC Conference on Computer and
  Communications Security (CCS)}}. ACM.
\newblock


\bibitem[\protect\citeauthoryear{Cui, Li, Liu, Wang, and K{\"u}hlewind}{Cui
  et~al\mbox{.}}{2017}]%
        {cui2017innovating}
\bibfield{author}{\bibinfo{person}{Yong Cui}, \bibinfo{person}{Tianxiang Li},
  \bibinfo{person}{Cong Liu}, \bibinfo{person}{Xingwei Wang}, {and}
  \bibinfo{person}{Mirja K{\"u}hlewind}.} \bibinfo{year}{2017}\natexlab{}.
\newblock \showarticletitle{Innovating transport with {QUIC}: Design approaches
  and research challenges}.
\newblock \bibinfo{journal}{\emph{IEEE Internet Computing}}
  \bibinfo{volume}{21}, \bibinfo{number}{2} (\bibinfo{year}{2017}),
  \bibinfo{pages}{72--76}.
\newblock


\bibitem[\protect\citeauthoryear{Delignat-Lavaud, Fournet, Kohlweiss,
  Protzenko, Rastogi, Swamy, Zanella-B{\'e}guelin, Bhargavan, Pan, and
  Zinzindohoue}{Delignat-Lavaud et~al\mbox{.}}{2017}]%
        {delignat2017implementing}
\bibfield{author}{\bibinfo{person}{Antoine Delignat-Lavaud},
  \bibinfo{person}{C{\'e}dric Fournet}, \bibinfo{person}{Markulf Kohlweiss},
  \bibinfo{person}{Jonathan Protzenko}, \bibinfo{person}{Aseem Rastogi},
  \bibinfo{person}{Nikhil Swamy}, \bibinfo{person}{Santiago
  Zanella-B{\'e}guelin}, \bibinfo{person}{Karthikeyan Bhargavan},
  \bibinfo{person}{Jianyang Pan}, {and} \bibinfo{person}{Jean~Karim
  Zinzindohoue}.} \bibinfo{year}{2017}\natexlab{}.
\newblock \showarticletitle{Implementing and proving the {TLS} 1.3 record
  layer}. In \bibinfo{booktitle}{\emph{Proc. IEEE Symposium on Security and
  Privacy (S\&P)}}. IEEE, \bibinfo{pages}{463--482}.
\newblock


\bibitem[\protect\citeauthoryear{Dittrich, Kenneally, et~al\mbox{.}}{Dittrich
  et~al\mbox{.}}{2012}]%
        {menloreport}
\bibfield{author}{\bibinfo{person}{David Dittrich}, \bibinfo{person}{Erin
  Kenneally}, {et~al\mbox{.}}} \bibinfo{year}{2012}\natexlab{}.
\newblock \showarticletitle{The Menlo Report: Ethical principles guiding
  information and communication technology research}.
\newblock \bibinfo{journal}{\emph{US Department of Homeland Security}}
  (\bibinfo{year}{2012}).
\newblock


\bibitem[\protect\citeauthoryear{Dowling, Fischlin, G{\"u}nther, and
  Stebila}{Dowling et~al\mbox{.}}{2015}]%
        {dowling2015cryptographic}
\bibfield{author}{\bibinfo{person}{Benjamin Dowling}, \bibinfo{person}{Marc
  Fischlin}, \bibinfo{person}{Felix G{\"u}nther}, {and}
  \bibinfo{person}{Douglas Stebila}.} \bibinfo{year}{2015}\natexlab{}.
\newblock \showarticletitle{A cryptographic analysis of the {TLS} 1.3 handshake
  protocol candidates}. In \bibinfo{booktitle}{\emph{Proc. ACM SIGSAC
  Conference on Computer and Communications Security (CCS)}}. ACM.
\newblock


\bibitem[\protect\citeauthoryear{Durumeric, Kasten, Bailey, and
  Halderman}{Durumeric et~al\mbox{.}}{2013a}]%
        {Durumeric2013}
\bibfield{author}{\bibinfo{person}{Z. Durumeric}, \bibinfo{person}{J. Kasten},
  \bibinfo{person}{M. Bailey}, {and} \bibinfo{person}{J.~A. Halderman}.}
  \bibinfo{year}{2013}\natexlab{a}.
\newblock \showarticletitle{Analysis of the {HTTPS} Certificate Ecosystem}. In
  \bibinfo{booktitle}{\emph{Proc. ACM Int. Measurement Conference (IMC)}}.
  \bibinfo{address}{Barcelona}.
\newblock


\bibitem[\protect\citeauthoryear{Durumeric, Wustrow, and Halderman}{Durumeric
  et~al\mbox{.}}{2013b}]%
        {durumeric2013zmap}
\bibfield{author}{\bibinfo{person}{Zakir Durumeric}, \bibinfo{person}{Eric
  Wustrow}, {and} \bibinfo{person}{J~Alex Halderman}.}
  \bibinfo{year}{2013}\natexlab{b}.
\newblock \showarticletitle{ZMap: Fast Internet-wide scanning and its security
  applications}. In \bibinfo{booktitle}{\emph{Proc. USENIX Security
  Symposium}}.
\newblock


\bibitem[\protect\citeauthoryear{Facebook}{Facebook}{2018}]%
        {fbFizzDeployment}
\bibfield{author}{\bibinfo{person}{Facebook}.} \bibinfo{year}{2018}\natexlab{}.
\newblock \bibinfo{title}{{Deploying TLS 1.3 at scale with Fizz, a performant
  open source TLS library}}.
\newblock \bibinfo{howpublished}{\url{https://code.fb.com/security/fizz/ }}.
\newblock


\bibitem[\protect\citeauthoryear{Facebook}{Facebook}{2019}]%
        {fbFizz}
\bibfield{author}{\bibinfo{person}{Facebook}.} \bibinfo{year}{2019}\natexlab{}.
\newblock \bibinfo{title}{{Fizz (Github)}}.
\newblock
  \bibinfo{howpublished}{\url{https://github.com/facebookincubator/fizz}}.
\newblock


\bibitem[\protect\citeauthoryear{Gamba, Rashed, Razaghpanah, Tapiador, and
  Vallina-Rodriguez}{Gamba et~al\mbox{.}}{2020}]%
        {gamba2020analysis}
\bibfield{author}{\bibinfo{person}{Julien Gamba}, \bibinfo{person}{Mohammed
  Rashed}, \bibinfo{person}{Abbas Razaghpanah}, \bibinfo{person}{Juan
  Tapiador}, {and} \bibinfo{person}{Narseo Vallina-Rodriguez}.}
  \bibinfo{year}{2020}\natexlab{}.
\newblock \showarticletitle{{An Analysis of Pre-installed Android Software}}.
  In \bibinfo{booktitle}{\emph{2020 IEEE Symposium on Security and Privacy
  (SP)}}. \bibinfo{publisher}{IEEE Computer Society}.
\newblock


\bibitem[\protect\citeauthoryear{Ghedini.}{Ghedini.}{[n.d.]}]%
        {yougettls13everyonedoes}
\bibfield{author}{\bibinfo{person}{Alessandro Ghedini.}}
  \bibinfo{year}{[n.d.]}\natexlab{}.
\newblock
\newblock


\bibitem[\protect\citeauthoryear{Gustafsson, Overier, Arlitt, and
  Carlsson}{Gustafsson et~al\mbox{.}}{2017}]%
        {firstlookct}
\bibfield{author}{\bibinfo{person}{Josef Gustafsson}, \bibinfo{person}{Gustaf
  Overier}, \bibinfo{person}{Martin Arlitt}, {and} \bibinfo{person}{Niklas
  Carlsson}.} \bibinfo{year}{2017}\natexlab{}.
\newblock \showarticletitle{{A First Look at the {CT} Landscape: {Certificate
  Transparency} Logs in Practice}}. In \bibinfo{booktitle}{\emph{Proc. Passive
  and Active Measurement (PAM)}}.
\newblock


\bibitem[\protect\citeauthoryear{Halvorson, Der, Foster, Savage, Saul, and
  Voelker}{Halvorson et~al\mbox{.}}{2015}]%
        {Halvorson2015}
\bibfield{author}{\bibinfo{person}{T. Halvorson}, \bibinfo{person}{M.~Fr. Der},
  \bibinfo{person}{I. Foster}, \bibinfo{person}{S. Savage},
  \bibinfo{person}{L.~K. Saul}, {and} \bibinfo{person}{G.~M Voelker}.}
  \bibinfo{year}{2015}\natexlab{}.
\newblock \showarticletitle{From .academy to .zone: An Analysis of the New
  {TLD} Land Rush}. In \bibinfo{booktitle}{\emph{Proc. ACM Int. Measurement
  Conference (IMC)}}. \bibinfo{address}{Tokyo}.
\newblock


\bibitem[\protect\citeauthoryear{Holz, Amann, Mehani, Wachs, and Kaafar}{Holz
  et~al\mbox{.}}{2016}]%
        {holz2015tls}
\bibfield{author}{\bibinfo{person}{Ralph Holz}, \bibinfo{person}{Johanna
  Amann}, \bibinfo{person}{Olivier Mehani}, \bibinfo{person}{Matthias Wachs},
  {and} \bibinfo{person}{Mohamed~Ali Kaafar}.} \bibinfo{year}{2016}\natexlab{}.
\newblock \showarticletitle{TLS in the wild: An Internet-wide analysis of
  TLS-based protocols for electronic communication}.
\newblock \bibinfo{journal}{\emph{Network and Distributed System Security
  Symposium (NDSS)}} (\bibinfo{year}{2016}).
\newblock


\bibitem[\protect\citeauthoryear{Holz, Braun, Kammenhuber, and Carle}{Holz
  et~al\mbox{.}}{2011}]%
        {holz11imc}
\bibfield{author}{\bibinfo{person}{Ralph Holz}, \bibinfo{person}{Lothar Braun},
  \bibinfo{person}{Nils Kammenhuber}, {and} \bibinfo{person}{Georg Carle}.}
  \bibinfo{year}{2011}\natexlab{}.
\newblock \showarticletitle{{The {SSL} Landscape: A Thorough Analysis of the
  {X.509 PKI} Using Active and Passive Measurements}}. In
  \bibinfo{booktitle}{\emph{Proc. ACM Int. Measurement Conference (IMC)}}.
\newblock


\bibitem[\protect\citeauthoryear{Jager, Schwenk, and Somorovsky}{Jager
  et~al\mbox{.}}{2015}]%
        {jager2015security}
\bibfield{author}{\bibinfo{person}{Tibor Jager}, \bibinfo{person}{J{\"o}rg
  Schwenk}, {and} \bibinfo{person}{Juraj Somorovsky}.}
  \bibinfo{year}{2015}\natexlab{}.
\newblock \showarticletitle{On the security of {TLS} 1.3 and {QUIC} against
  weaknesses in {PKCS\#} 1 v1. 5 encryption}. In
  \bibinfo{booktitle}{\emph{Proc. ACM SIGSAC Conference on Computer and
  Communications Security (CCS)}}. ACM.
\newblock


\bibitem[\protect\citeauthoryear{Kakhki, Jero, Choffnes, Nita-Rotaru, and
  Mislove}{Kakhki et~al\mbox{.}}{2017}]%
        {kakhki2017taking}
\bibfield{author}{\bibinfo{person}{Arash~Molavi Kakhki},
  \bibinfo{person}{Samuel Jero}, \bibinfo{person}{David Choffnes},
  \bibinfo{person}{Cristina Nita-Rotaru}, {and} \bibinfo{person}{Alan
  Mislove}.} \bibinfo{year}{2017}\natexlab{}.
\newblock \showarticletitle{Taking a long look at {QUIC}: an approach for
  rigorous evaluation of rapidly evolving transport protocols}. In
  \bibinfo{booktitle}{\emph{Proc. ACM Int. Measurement Conference (IMC)}}. ACM.
\newblock


\bibitem[\protect\citeauthoryear{Kobeissi}{Kobeissi}{2018}]%
        {kobeissi2018formal}
\bibfield{author}{\bibinfo{person}{Nadim Kobeissi}.}
  \bibinfo{year}{2018}\natexlab{}.
\newblock \emph{\bibinfo{title}{Formal Verification for Real-World
  Cryptographic Protocols and Implementations}}.
\newblock \bibinfo{thesistype}{Ph.D. Dissertation}. \bibinfo{school}{INRIA
  Paris; Ecole Normale Sup{\'e}rieure de Paris-ENS Paris}.
\newblock


\bibitem[\protect\citeauthoryear{Kotzias, Razaghpanah, Amann, Paterson,
  Vallina-Rodriguez, and Caballero}{Kotzias et~al\mbox{.}}{2018}]%
        {kotzias2018coming}
\bibfield{author}{\bibinfo{person}{Platon Kotzias}, \bibinfo{person}{Abbas
  Razaghpanah}, \bibinfo{person}{Johanna Amann}, \bibinfo{person}{Kenneth~G
  Paterson}, \bibinfo{person}{Narseo Vallina-Rodriguez}, {and}
  \bibinfo{person}{Juan Caballero}.} \bibinfo{year}{2018}\natexlab{}.
\newblock \showarticletitle{Coming of Age: A Longitudinal Study of {TLS}
  Deployment}. In \bibinfo{booktitle}{\emph{Proc. ACM Int. Measurement
  Conference (IMC)}}.
\newblock


\bibitem[\protect\citeauthoryear{Krawczyk and Wee}{Krawczyk and Wee}{2016}]%
        {krawczyk2016optls}
\bibfield{author}{\bibinfo{person}{Hugo Krawczyk} {and}
  \bibinfo{person}{Hoeteck Wee}.} \bibinfo{year}{2016}\natexlab{}.
\newblock \showarticletitle{The {OPTLS} protocol and {TLS} 1.3}. In
  \bibinfo{booktitle}{\emph{Proc. IEEE European Symposium on Security and
  Privacy (EuroS\&P)}}. IEEE.
\newblock


\bibitem[\protect\citeauthoryear{Langley, Riddoch, Wilk, Vicente, Krasic,
  Zhang, Yang, Kouranov, Swett, Iyengar, et~al\mbox{.}}{Langley
  et~al\mbox{.}}{2017}]%
        {langley2017quic}
\bibfield{author}{\bibinfo{person}{Adam Langley}, \bibinfo{person}{Alistair
  Riddoch}, \bibinfo{person}{Alyssa Wilk}, \bibinfo{person}{Antonio Vicente},
  \bibinfo{person}{Charles Krasic}, \bibinfo{person}{Dan Zhang},
  \bibinfo{person}{Fan Yang}, \bibinfo{person}{Fedor Kouranov},
  \bibinfo{person}{Ian Swett}, \bibinfo{person}{Janardhan Iyengar},
  {et~al\mbox{.}}} \bibinfo{year}{2017}\natexlab{}.
\newblock \showarticletitle{The {QUIC} transport protocol: Design and
  internet-scale deployment}. In \bibinfo{booktitle}{\emph{Proceedings of the
  Conference of the ACM Special Interest Group on Data Communication}}. ACM,
  \bibinfo{pages}{183--196}.
\newblock


\bibitem[\protect\citeauthoryear{Laurie, Langley, and Kasper}{Laurie
  et~al\mbox{.}}{2013}]%
        {RFC6962}
\bibfield{author}{\bibinfo{person}{B. Laurie}, \bibinfo{person}{A. Langley},
  {and} \bibinfo{person}{E. Kasper}.} \bibinfo{year}{2013}\natexlab{}.
\newblock \bibinfo{booktitle}{\emph{{Certificate Transparency}}}.
\newblock \bibinfo{type}{RFC} 6962. \bibinfo{institution}{IETF}.
\newblock
\urldef\tempurl%
\url{http://tools.ietf.org/rfc/rfc6962.txt}
\showURL{%
\tempurl}


\bibitem[\protect\citeauthoryear{Lychev, Jero, Boldyreva, and
  Nita-Rotaru}{Lychev et~al\mbox{.}}{2015}]%
        {lychev2015secure}
\bibfield{author}{\bibinfo{person}{Robert Lychev}, \bibinfo{person}{Samuel
  Jero}, \bibinfo{person}{Alexandra Boldyreva}, {and} \bibinfo{person}{Cristina
  Nita-Rotaru}.} \bibinfo{year}{2015}\natexlab{}.
\newblock \showarticletitle{How secure and quick is {QUIC}? Provable security
  and performance analyses}. In \bibinfo{booktitle}{\emph{2015 IEEE Symposium
  on Security and Privacy}}. IEEE, \bibinfo{pages}{214--231}.
\newblock


\bibitem[\protect\citeauthoryear{M{\"o}ller, Duong, and Kotowicz}{M{\"o}ller
  et~al\mbox{.}}{2014}]%
        {moller2014poodle}
\bibfield{author}{\bibinfo{person}{Bodo M{\"o}ller}, \bibinfo{person}{Thai
  Duong}, {and} \bibinfo{person}{Krzysztof Kotowicz}.}
  \bibinfo{year}{2014}\natexlab{}.
\newblock \showarticletitle{This {POODLE} bites: exploiting the {SSL} 3.0
  fallback}.
\newblock \bibinfo{journal}{\emph{Security Advisory}} (\bibinfo{year}{2014}).
\newblock


\bibitem[\protect\citeauthoryear{{Nick Sullivan}}{{Nick Sullivan}}{2016}]%
        {tls13}
\bibfield{author}{\bibinfo{person}{{Nick Sullivan}}.}
  \bibinfo{year}{2016}\natexlab{}.
\newblock \bibinfo{title}{{Introducing TLS 1.3}}.
\newblock
\newblock
\newblock
\shownote{\url{https://blog.cloudflare.com/introducing-tls-1-3/}.}


\bibitem[\protect\citeauthoryear{{Nick Sullivan}}{{Nick Sullivan}}{2017}]%
        {tls13-0rttCloudflare}
\bibfield{author}{\bibinfo{person}{{Nick Sullivan}}.}
  \bibinfo{year}{2017}\natexlab{}.
\newblock \bibinfo{title}{{Introducing Zero Round Trip Time Resumption
  (0-RTT)}}.
\newblock
\newblock
\newblock
\shownote{\url{https://blog.cloudflare.com/introducing-0-rtt/}.}


\bibitem[\protect\citeauthoryear{Partridge and Allman}{Partridge and
  Allman}{2016}]%
        {PA16}
\bibfield{author}{\bibinfo{person}{Craig Partridge} {and} \bibinfo{person}{Mark
  Allman}.} \bibinfo{year}{2016}\natexlab{}.
\newblock \showarticletitle{{Addressing Ethical Considerations in Network
  Measurement Papers}}.
\newblock \bibinfo{journal}{\emph{Commun. ACM}} \bibinfo{volume}{59},
  \bibinfo{number}{10} (\bibinfo{date}{Oct.} \bibinfo{year}{2016}).
\newblock


\bibitem[\protect\citeauthoryear{Razaghpanah, Niaki, Vallina-Rodriguez,
  Sundaresan, Amann, and Gill}{Razaghpanah et~al\mbox{.}}{2017}]%
        {razaghpanah2017studying}
\bibfield{author}{\bibinfo{person}{Abbas Razaghpanah},
  \bibinfo{person}{Arian~Akhavan Niaki}, \bibinfo{person}{Narseo
  Vallina-Rodriguez}, \bibinfo{person}{Srikanth Sundaresan},
  \bibinfo{person}{Johanna Amann}, {and} \bibinfo{person}{Phillipa Gill}.}
  \bibinfo{year}{2017}\natexlab{}.
\newblock \showarticletitle{Studying {TLS} usage in Android apps}. In
  \bibinfo{booktitle}{\emph{Proc. ACM Int. Conference on emerging Networking
  EXperiments and Technologies (CoNEXT)}}.
\newblock


\bibitem[\protect\citeauthoryear{Razaghpanah, Vallina-Rodriguez, Sundaresan,
  Kreibich, Gill, Allman, and Paxson}{Razaghpanah et~al\mbox{.}}{2015}]%
        {razaghpanah2015haystack}
\bibfield{author}{\bibinfo{person}{Abbas Razaghpanah}, \bibinfo{person}{Narseo
  Vallina-Rodriguez}, \bibinfo{person}{Srikanth Sundaresan},
  \bibinfo{person}{Christian Kreibich}, \bibinfo{person}{Phillipa Gill},
  \bibinfo{person}{Mark Allman}, {and} \bibinfo{person}{Vern Paxson}.}
  \bibinfo{year}{2015}\natexlab{}.
\newblock \showarticletitle{Haystack: A multi-purpose mobile vantage point in
  user space}.
\newblock \bibinfo{journal}{\emph{arXiv preprint arXiv:1510.01419}}
  (\bibinfo{year}{2015}).
\newblock


\bibitem[\protect\citeauthoryear{Rescolla}{Rescolla}{2016}]%
        {tls13-draft16}
\bibfield{author}{\bibinfo{person}{Eric Rescolla}.}
  \bibinfo{year}{2016}\natexlab{}.
\newblock \bibinfo{title}{{The {Transport Layer Security (TLS)} Protocol
  Version 1.3 - Draft 16}}.
\newblock
\newblock
\urldef\tempurl%
\url{https://tools.ietf.org/html/draft-ietf-tls-tls13-16}
\showURL{%
\tempurl}


\bibitem[\protect\citeauthoryear{Rescolla}{Rescolla}{2017}]%
        {tls13-draft22}
\bibfield{author}{\bibinfo{person}{Eric Rescolla}.}
  \bibinfo{year}{2017}\natexlab{}.
\newblock \bibinfo{title}{{The {Transport Layer Security (TLS)} Protocol
  Version 1.3 - Draft 16}}.
\newblock
\newblock
\urldef\tempurl%
\url{https://tools.ietf.org/html/draft-ietf-tls-tls13-22}
\showURL{%
\tempurl}


\bibitem[\protect\citeauthoryear{Rescolla}{Rescolla}{2018}]%
        {rfc8446}
\bibfield{author}{\bibinfo{person}{Eric Rescolla}.}
  \bibinfo{year}{2018}\natexlab{}.
\newblock \bibinfo{title}{{The {Transport Layer Security (TLS)} Protocol
  Version 1.3}}.
\newblock \bibinfo{howpublished}{RFC 8446 (Historic)}.
\newblock
\showISSN{2070-1721}
\urldef\tempurl%
\url{https://tools.ietf.org/html/rfc8446}
\showURL{%
\tempurl}
\newblock
\shownote{RFC 8446.}


\bibitem[\protect\citeauthoryear{R{\"u}th, Poese, Dietzel, and
  Hohlfeld}{R{\"u}th et~al\mbox{.}}{2018}]%
        {ruth2018first}
\bibfield{author}{\bibinfo{person}{Jan R{\"u}th}, \bibinfo{person}{Ingmar
  Poese}, \bibinfo{person}{Christoph Dietzel}, {and} \bibinfo{person}{Oliver
  Hohlfeld}.} \bibinfo{year}{2018}\natexlab{}.
\newblock \showarticletitle{A First Look at QUIC in the Wild}. In
  \bibinfo{booktitle}{\emph{International Conference on Passive and Active
  Network Measurement}}. Springer, \bibinfo{pages}{255--268}.
\newblock


\bibitem[\protect\citeauthoryear{Ryan}{Ryan}{2014}]%
        {ctemail}
\bibfield{author}{\bibinfo{person}{Mark~D. Ryan}.}
  \bibinfo{year}{2014}\natexlab{}.
\newblock \showarticletitle{{Enhanced Certificate Transparency and End-to-End
  Encrypted Mail}}. In \bibinfo{booktitle}{\emph{Network and Distributed System
  Security Symposium (NDSS)}}.
\newblock


\bibitem[\protect\citeauthoryear{Scheitle, Chung, Amann, Gasser, Brent, Carle,
  Holz, Hiller, Naab, van Rijswijk-Deij, et~al\mbox{.}}{Scheitle
  et~al\mbox{.}}{2018a}]%
        {scheitle2018measuring}
\bibfield{author}{\bibinfo{person}{Quirin Scheitle}, \bibinfo{person}{Taejoong
  Chung}, \bibinfo{person}{Johanna Amann}, \bibinfo{person}{Oliver Gasser},
  \bibinfo{person}{Lexi Brent}, \bibinfo{person}{Georg Carle},
  \bibinfo{person}{Ralph Holz}, \bibinfo{person}{Jens Hiller},
  \bibinfo{person}{Johannes Naab}, \bibinfo{person}{Roland van Rijswijk-Deij},
  {et~al\mbox{.}}} \bibinfo{year}{2018}\natexlab{a}.
\newblock \showarticletitle{Measuring Adoption of Security Additions to the
  {HTTPS} Ecosystem}. In \bibinfo{booktitle}{\emph{Proceedings of the Applied
  Networking Research Workshop}}. ACM.
\newblock


\bibitem[\protect\citeauthoryear{Scheitle, Gasser, Nolte, Amann, Brent, Carle,
  Holz, Schmidt, and W{\"a}hlisch}{Scheitle et~al\mbox{.}}{2018b}]%
        {scheitle2018rise}
\bibfield{author}{\bibinfo{person}{Quirin Scheitle}, \bibinfo{person}{Oliver
  Gasser}, \bibinfo{person}{Theodor Nolte}, \bibinfo{person}{Johanna Amann},
  \bibinfo{person}{Lexi Brent}, \bibinfo{person}{Georg Carle},
  \bibinfo{person}{Ralph Holz}, \bibinfo{person}{Thomas~C Schmidt}, {and}
  \bibinfo{person}{Matthias W{\"a}hlisch}.} \bibinfo{year}{2018}\natexlab{b}.
\newblock \showarticletitle{The Rise of {Certificate Transparency} and Its
  Implications on the Internet Ecosystem}. In
  \bibinfo{booktitle}{\emph{Proceedings of the Internet Measurement Conference
  2018}}. ACM, \bibinfo{pages}{343--349}.
\newblock


\bibitem[\protect\citeauthoryear{Scheitle, Hohlfeld, Gamba, Jelten, Zimmermann,
  Strowes, and Vallina-Rodriguez}{Scheitle et~al\mbox{.}}{2018c}]%
        {Scheitle2018}
\bibfield{author}{\bibinfo{person}{Quirin Scheitle}, \bibinfo{person}{Oliver
  Hohlfeld}, \bibinfo{person}{Julien Gamba}, \bibinfo{person}{Jonas Jelten},
  \bibinfo{person}{Torsten Zimmermann}, \bibinfo{person}{Stephen~D. Strowes},
  {and} \bibinfo{person}{Narseo Vallina-Rodriguez}.}
  \bibinfo{year}{2018}\natexlab{c}.
\newblock \showarticletitle{{A Long Way to the Top: Significance, Structure,
  and Stability of Internet Top Lists}}. In \bibinfo{booktitle}{\emph{Proc. ACM
  Int. Measurement Conference (IMC)}}. ACM.
\newblock


\bibitem[\protect\citeauthoryear{Thomson}{Thomson}{2018}]%
        {RFC8449}
\bibfield{author}{\bibinfo{person}{M. Thomson}.}
  \bibinfo{year}{2018}\natexlab{}.
\newblock \bibinfo{booktitle}{\emph{{Record Size Limit Extension for TLS}}}.
\newblock \bibinfo{type}{RFC} 8449. \bibinfo{institution}{IETF}.
\newblock
\urldef\tempurl%
\url{http://tools.ietf.org/rfc/rfc8449.txt}
\showURL{%
\tempurl}


\bibitem[\protect\citeauthoryear{VanderSloot, Amann, Bernhard, Durumeric,
  Bailey, and Halderman}{VanderSloot et~al\mbox{.}}{2016}]%
        {vandersloot2016towards}
\bibfield{author}{\bibinfo{person}{B. VanderSloot}, \bibinfo{person}{J. Amann},
  \bibinfo{person}{M. Bernhard}, \bibinfo{person}{Z. Durumeric},
  \bibinfo{person}{M. Bailey}, {and} \bibinfo{person}{J.A. Halderman}.}
  \bibinfo{year}{2016}\natexlab{}.
\newblock \showarticletitle{Towards a Complete View of the Certificate
  Ecosystem}. In \bibinfo{booktitle}{\emph{Proc. ACM Int. Measurement
  Conference (IMC)}}.
\newblock


\bibitem[\protect\citeauthoryear{Yilek, Rescorla, Shacham, Enright, and
  Savage}{Yilek et~al\mbox{.}}{2009}]%
        {debiankeys}
\bibfield{author}{\bibinfo{person}{Scott Yilek}, \bibinfo{person}{Eric
  Rescorla}, \bibinfo{person}{Hovav Shacham}, \bibinfo{person}{Brandon
  Enright}, {and} \bibinfo{person}{Stefan Savage}.}
  \bibinfo{year}{2009}\natexlab{}.
\newblock \showarticletitle{{When Private Keys Are Public: Results from the
  2008 {Debian OpenSSL} Vulnerability}}. In \bibinfo{booktitle}{\emph{Proc. ACM
  Int. Measurement Conference (IMC)}}.
\newblock


\bibitem[\protect\citeauthoryear{Zhang, Choffnes, Levin, Dumitras, Mislove,
  Schulman, and Wilson}{Zhang et~al\mbox{.}}{2014}]%
        {wake_of_heartbleed}
\bibfield{author}{\bibinfo{person}{L. Zhang}, \bibinfo{person}{D. Choffnes},
  \bibinfo{person}{D. Levin}, \bibinfo{person}{T. Dumitras},
  \bibinfo{person}{A. Mislove}, \bibinfo{person}{A. Schulman}, {and}
  \bibinfo{person}{C. Wilson}.} \bibinfo{year}{2014}\natexlab{}.
\newblock \showarticletitle{Analysis of {SSL} Certificate Reissues and
  Revocations in the Wake of {H}eartbleed}. In \bibinfo{booktitle}{\emph{Proc.
  ACM Int. Measurement Conference (IMC)}}.
\newblock


\end{thebibliography}
}
\appendix

\section{Ethical considerations}
\label{sec:appendix_ethics}

Our study involves the passive collection of network traffic from real users 
and active network scans. We follow the principles of informed
consent~\cite{menloreport} and best practices~\cite{PA16}: 
we avoid the collection of any personal or
sensitive data, such as client IP addresses or traffic payloads, and we try to
avoid causing any harm to online servers during our active scans. 
Below we discuss details specific to each tool.

\subsection{Passive Data collection}

The passive data collection effort performed by the ICSI SSL Notary was cleared
by the respective responsible parties at each contributing institution before
they began contributing. Note that the ICSI SSL Notary specifically excludes or
anonymizes sensitive information, such as client IP addresses. In more detail,
client IP addresses are combined with the server IP address and SNI as well as
a site-specific, secret, hash unknown to ICSI. The resulting string is hashed.
This allows the dection of when the same client connected to the same IP address
(\eg to evaluate the effectiveness of session resumption), without enabling the
tracking of a client while it accesses different servers. It also means that ICSI
data does not contain any information of how many users are active at a specific
site. While the Notary records server-sent certificates the notary does not record
client-certificates if they are present in the handshake. The Notary only records
handshake information that is sent in the clear.

Passive data collection at the Australian hosting institution was reviewed and
approved by the responsible Human Ethics board. The data collection follows the same
anonymization principles as the ICSI Notary.

\subsection{Active Scans}

We took precautions to minimize the impact of our scans, following established
practices as, for instance, described in~\cite{durumeric2013zmap}. In
particular, we maintain a blacklist to avoid scanning systems that have in the
past indicated to us that they do not wish to be scanned. Our abuse email
address is published in the WHOIS and all abuse emails are forwarded to us by
our IT department. We received one abuse email sent by a blocklist provider;
our scanner was whitelisted when we explained our work. Our scanning activity
was also reviewed by the Human Ethics board of our hosting institution; it was
found that we do not collect personally identifiable information and hence need
not undergo a Human Ethics approval process. We assess the impact of our
scans in terms of potential harm to other systems and human beings, as proposed
by the Menlo report~\cite{menloreport}. We use a relatively low scanning rate
to minimize any impact and respond immediately to complaints.

\subsection{Lumen Privacy Monitor}
\label{appendix:lumenethics}

Lumen's view of real-world mobile data collected from end-user devices 
raises ethical issues.  We address these in two ways:

\noindent \textbf{Informed Consent.\xspace}
Lumen follows the principles of informed consent as indicated by the Menlo
Report~\cite{menloreport} and avoids the collection of any personal or sensitive 
data. Users must explicitly grant permission to Lumen to inspect the traffic 
and the app
requires users to opt-in a second time to install a CA certificate to inspect
encrypted traffic. Furthermore, the user can disable traffic interception and
uninstall the app at any time. The privacy policy of the app is available in
Google Play as well as in the project's website:
\url{https://haystack.mobi/privacy.html}.

\noindent \textbf{Data Collection Strategy. \xspace}
Lumen runs on the user's device. It allows Lumen to confine the bulk of the data
processing to the device itself. Lumen only collects and uploads to the
project servers' anonymized summary statistics. Mobile app traffic 
flows are mapped to the
app generating them, and not to a user identity. For example, we collect flow
metadata like  
TLS Client Hello and Server Hello records, HTTP User Agent Field, 
byte counts, the destination IP address and
the remote TCP port number, the
package name and version of the app making the connection, and the
OS version of Android running on the device. 

The data is uploaded following reasonable security mechanisms
(\ie use of encryption).
To further protect user privacy, Lumen also ignores flows generated by
applications which may potentially deanonymize a user. Examples of
such applications are mobile browsers such as the \textit{Android default
  browser} or \textit{Google  Chrome}. The type of traffic
generated by these apps is highly dependent of user actions, which not only
makes deanonymizing users easier, but also beats our purpose of understanding
the way that mobile apps work due to developer decisions.
The team behind Lumen follows ethical protocols, which were developed 
in consultation with their Institutional Review Board (IRB) ---it is considered
as a non-human subject research effort due to the anonimization process---
before starting any data collection.

\end{document}